\documentclass[english]{article}
\usepackage[utf8]{inputenc}
\usepackage{lmodern}

\usepackage[T1]{fontenc}
\usepackage{color}
\usepackage{mathrsfs}
\usepackage{amsmath}
\usepackage{amssymb}
\usepackage{colortbl}
\usepackage{booktabs}
\usepackage{multirow}

\makeatletter

\usepackage{mathrsfs}   
\usepackage{slashed}     
\usepackage{bbold}  
\usepackage{url}
\usepackage{graphicx}
\usepackage[colorlinks=true,linkcolor=redLinks,citecolor=greenLinks,urlcolor=redLinks, pdfborder={0 0 1}]{hyperref}
\usepackage{xcolor}
\usepackage{framed}
\usepackage[numbers,sort&compress]{natbib}
\usepackage{amsmath}

\allowdisplaybreaks

\definecolor{greenLinks}{rgb}{0, 0.6, 0} 
\definecolor{blueLinks}{rgb}{0, 0, 0.6}
\definecolor{redLinks}{rgb}{0.6, 0, 0}

\usepackage{multirow}
\newcommand{\ol}{16\pi^2\,}
\newcommand{\lag}{\mathscr{L}}

\newcommand{\afive}[2]{\big(a^{(5)}_{#1}\big)_{#2}}
\newcommand{\rfive}[2]{\big(r^{(5)}_{#1}\big)_{#2}}
\newcommand{\afivep}[2]{\big(a^{(5)~\prime}_{#1}\big)_{#2}}
\newcommand{\rfivep}[2]{\big(r^{(5)~\prime}_{#1}\big)_{#2}}
\newcommand{\afived}[2]{\big(\dot{a}^{(5)}_{#1}\big)_{#2}}
\newcommand{\asix}[2]{\big(a^{(6)}_{#1}\big)_{#2}}
\newcommand{\rsix}[2]{\big(r^{(6)}_{#1}\big)_{#2}}
\newcommand{\asixp}[2]{\big(a^{(6)~\prime}_{#1}\big)_{#2}}
\newcommand{\asixd}[2]{\big(\dot{a}^{(6)}_{#1}\big)_{#2}}
\newcommand{\rsixp}[2]{\big(r^{(6)~\prime}_{#1}\big)_{#2}}

\newcommand{\tr}[1]{\mathrm{Tr}[#1]}

\definecolor{RFcolor}{rgb}{0.48,0,0.76}

\newcommand{\RF}[1]{{\leavevmode\color{violet}{#1}}}

\makeatother

\usepackage{babel}
\begin{document}
\title{Renormalization of general Effective Field Theories: 
\\Formalism and renormalization of bosonic operators}
\author{Renato M. Fonseca${}^a$, Pablo Olgoso${}^{b,c}$ and Jos\'e Santiago${}^a$}

\maketitle

\begin{center}
{\Large{}\vspace{-0.5cm}}
${}^a$Departamento de F\'isica Te\'orica y del Cosmos,
Universidad de Granada, \\
Campus de Fuentenueva, E--18071 Granada,
Spain
~\\
${}^b$ Dipartimento di Fisica e Astronomia, Universit\`a di Padova, Via F. Marzolo 8, 35131 Padova, Italy
~\\
${}^c$ Istituto Nazionale di Fisica Nucleare, Sezione di Padova, Via F. Marzolo 8, 35131 Padova, Italy
~\\
Emails: renatofonseca@ugr.es, pablo.olgosoruiz@unipd.it, jsantiago@ugr.es
\par\end{center}

\begin{abstract} We describe the most general local, Lorentz-invariant, effective field theory of scalars, fermions and gauge bosons up to mass dimension 6. We first obtain both a Green and a physical basis for such an effective theory, together with the on-shell reduction of the former to the latter. We then proceed to compute the renormalization group equations for the bosonic operators of this general effective theory at one-loop order.
\end{abstract}

\section{Introduction}

Effective field theory (EFT) techniques are essential in the study of physical systems with disparate scales. The large logarithms appearing in quantum calculations of observables in these systems can easily destabilize perturbation theory unless the renormalization group equations (RGEs) are used to re-sum them. The calculation of the beta functions of a specific EFT is a well known procedure and has already been fully automated, at the one loop order, thanks to tools like \texttt{MatchMakerEFT}~\cite{Carmona:2021xtq} or \texttt{Matchete}~\cite{Fuentes-Martin:2022jrf} (see recent progress towards the automation of two loop calculations in~\cite{Fuentes-Martin:2023ljp,Fuentes-Martin:2024agf}) and one-loop calculations for some of the most phenomenologically relevant EFTs have been performed in the literature, including the LEFT~\cite{Jenkins:2017dyc} and SMEFT~\cite{Jenkins:2013zja,Jenkins:2013wua,Alonso:2013hga} up to dimension 6, the SMEFT partially at dimension 8~\cite{Chala:2021pll,AccettulliHuber:2021uoa,DasBakshi:2022mwk,Helset:2022pde,Assi:2023zid,Boughezal:2024zqa,Bakshi:2024wzz}, or the ALP-LEFT and ALP-SMEFT~\cite{Chala:2020wvs,Bauer:2020jbp,Bonilla:2021ufe,Bresciani:2024shu} and even beyond~\cite{DasBakshi:2023lca}.~\footnote{Some partial two-loop calculations of the most relevant RGEs are already available in the literature, including~\cite{Bern:2020ikv,Jenkins:2023bls,DiNoi:2024ajj,Born:2024mgz}.} Still, the corresponding calculation is always tedious, even with automated tools, and long, and has to be repeated for every EFT of interest. Following the pioneering work of Refs.~\cite{Machacek:1983fi,Machacek:1983tz,Machacek:1984zw}
 (see also~\cite{Martin:1993zk,Luo:2002ti}), we propose here to define a completely general EFT, with no restriction of field content or symmetry group, up to mass dimension 6. In order to allow for an off-shell renormalization (and, eventually, finite matching) we define a Green's basis and a physical basis, paying particular attention to the symmetry properties of the different operators and their corresponding Wilson coefficients (WCs). 
 We provide the complete reduction of the Green's basis to the physical one using field redefinitions, that correctly incorporate non-linear terms in the reduction~\cite{Criado:2018sdb}. These terms are irrelevant for the one-loop renormalization of the theory but are important for finite matching and higher-loop renormalization. We then proceed to renormalize the bosonic operators of our general EFT at one loop order, including both bosonic and fermionic operators as boundary condition in the UV. The corresponding counterterms allow us to compute the beta functions of all bosonic operators in the physical basis. With this calculation one can obtain the one-loop beta functions of the bosonic operators of any EFT up to mass dimension 6 by means of a straight-forward group-theoretical calculation. The renormalization of the fermionic operators, and the corresponding beta functions, will be provided in a companion article~\cite{companion_fermions}. 

 The rest of the article is organized as follows. We describe the most general effective Lagrangian up to mass dimension 6 in Section~\ref{sec_general_EFT}, with special attention to the symmetry of the different operators and the fermionic evanescent operators. We provide the reduction of the Green's basis to the physical one in Section~\ref{sec_reduction}. Section~\ref{sec_beta_functions} is devoted to the calculation of the beta functions and contains one of the most relevant results of our work, the bosonic beta functions of our general EFT. We end with our conclusions in Section~\ref{sec_conclusions} and leave some technical details to the appendices, with Appendix~\ref{appendix_RGE} devoted to the general calculation of beta functions and~\ref{matching:green} to the result of the one-loop divergences in the Green's basis.

\section{The general effective Lagrangian to mass dimension 6 \label{sec_general_EFT}}

Let us introduce our notation for the fields and symmetries defining our general EFT. We will consider a general gauge group $G$, which can be any compact Lie group, with structure constants $f^{ABC}$ and gauge bosons $V_\mu^A$. The field strength tensor reads
\begin{equation}
    F^A_{\mu\nu}=\partial_\mu V^A_\nu - \partial_\nu V^A_\mu + g_A f^{ABC} V^B_\mu V^C_\nu,
\end{equation}
and its covariant derivative
\begin{equation}
D_\rho F^A_{\mu\nu}\equiv (D_\rho F_{\mu\nu})^A = \partial_\rho F^A_{\mu\nu} + g_A f^{ABC} V^B_\rho F^C_{\mu\nu}.    
\end{equation}
All fermions --- taken to be left-handed (LH) --- can be collected in a single vector $\psi_i$ which transforms under some representation  of $G$, which is in general reducible. Likewise, all scalars $\phi_i$ are considered real and grouped in a single (potentially reducible) representation of $G$. Our convention for the covariant derivatives applied to these fields reads
\begin{align}
    D_\mu \psi_i &= \partial_\mu \psi_i - \mathrm{i} g_A t^A_{ij} V^A_\mu \psi_j, \\
    D_\mu \phi_a &= \partial_\mu \phi_a - \mathrm{i} g_A \theta^A_{ab} V^A_\mu \phi_b.
\end{align}
We adopt the convention that, unless otherwise indicated, repeated indices are to be summed over. Given that gauge transformations preserve the norm of fields, both the $t^A$ and the $\theta^A$ matrices must be hermitian. Additionally, since we are working with real scalars,  the $\theta^A$ must be purely imaginary and therefore anti-symmetric.

The renormalizable Lagrangian reads
\begin{align}
    \lag_{d\leq 4} =& -\frac{1}{4} (a_{KF})_{AB} F^A_{\mu\nu} F^{B\, \mu\nu}+ \frac{1}{2} (a_{K\phi})_{ab} D_\mu \phi_a D^\mu \phi_b
    + (a_{K\psi})_{ij} \bar{\psi}_i \mathrm{i} \slashed{D} \psi_j
    -\frac{1}{2} \Big[ (m_\psi)_{ij} \psi^T_i C \psi_j  + \mathrm{h.c.}\Big] 
\nonumber \\
&    -\frac{1}{2} (m_\phi^2)_{ab} \phi_a \phi_b 
-\eta_a \phi_a -\frac{1}{2} \Big[ Y_{ija} \psi^T_i C \psi_j  + \mathrm{h.c.}\Big] \phi_a
    -\frac{\kappa_{abc}}{3!} \phi_a \phi_b \phi_c
    -\frac{\lambda_{abcd}}{4!} \phi_a \phi_b \phi_c \phi_d,\label{L1}
\end{align}
where $C$ is the charge conjugation matrix, we have allowed arbitrary kinetic terms as needed to recover non-trivial wave-function renormalization, and have allowed for a tadpole term, again needed for renormalization, which is possible only for singlets under the gauge symmetry. 
The WCs in this as well as other Lagrangians which
we will encounter later, are not fully generic for the following reasons:
\begin{enumerate}
\item The symmetry properties of a tensor under permutations
of its indices are transmitted to other tensors to which it is contracted.
For example, if $\mathcal{O}_{ij}$ is symmetric then in its contraction
$w_{ij}\mathcal{O}_{ij}$ with a generic tensor $w$ the anti-symmetric
part of the latter tensor is projected out, and for all practical
considerations we can take $w$ to be symmetric as well. In a Lagrangian,
products of fields play the role of $\mathcal{O}$ in this example,
setting the symmetry of the Wilson coefficient $w$ which contracts
with it. This principle remains true even for more complicated symmetries
which we discuss in this work.
\item The Lagrangian as a whole must be self-conjugate.
\end{enumerate}
As such, in Eq. (\ref{L1}) one must take into account the following symmetries:
\begin{align}
    (a_{KF})_{AB}&=(a_{KF})_{BA},\quad
    (a_{K\phi})_{ab}=(a_{K\phi})_{ba},\quad
    (a_{K\psi})_{ij}=(a_{K\psi})^\ast_{ji},\\
    (m_\psi)_{ij}&=(m_\psi)_{ji},\quad
    (m_\phi^2)_{ab}=(m_\phi^2)_{ba}, \quad
    Y_{ija}=Y_{jia}, \\
    \kappa_{abc}&=\mbox{fully symmetric}, \quad
    \lambda_{abcd}=\mbox{fully symmetric}.
\end{align}
All couplings are real, except for the ones involving fermions, namely $a_{K\psi}$, $m_\psi$ and $Y$.

In the following we are going to list the Green's basis for this generic theory at mass dimensions 5 and 6, obtained in part with the help of the \texttt{Sym2Int} package \cite{Fonseca:2017lem,Fonseca:2019yya}. We will use the following notation for the general effective Lagrangian. Operators of mass dimension $d$ in the Green's basis and their Wilson coefficients will be denoted by 
\begin{equation}
\lag_d^{\mathrm{Green}}=\lag_d^{\mathrm{phys}}
+\lag_d^{\mathrm{red}},
\end{equation}
with the operators in the physical and redundant bases, reading, respectively,
\begin{align}
\lag_d^{\mathrm{phys}} &=
\sum_x (a^{(d)}_{x})_{ij\ldots} (\mathcal{O}^{(d)}_x)_{ij\ldots}, \\
\lag_d^{\mathrm{red}} &=
\sum_x (r^{(d)}_{x})_{ij\ldots} (\mathcal{R}^{(d)}_x)_{ij\ldots},
\end{align}
where $ij\ldots$ denote the corresponding indices.

At mass dimension 5 we have,
\begin{align}
    \lag_5^{\mathrm{phys}} &=
    \bigg[ \frac{1}{2} \afive{\psi F}{Aij} \psi_i^T C \sigma^{\mu \nu} \psi_j F^A_{\mu\nu} 
    +\frac{1}{4} \afive{\psi \phi^2}{ijab}  \psi_i^T C \psi_j \phi_a \phi_b  
    + \mathrm{h.c.} \bigg]
    \nonumber \\
    &+\frac{1}{2} \afive{\phi F}{ABa} F^{A\, \mu\nu} F^B_{\mu\nu} \phi_a
    +\frac{1}{2} \afive{\phi \widetilde{F}}{ABa} F^{A\,\mu\nu} 
    \widetilde{F}^B_{\mu\nu} \phi_a 
    + \frac{1}{5!}\afive{\phi}{abcde} \phi_a \phi_b \phi_c \phi_d \phi_e, \label{L5phys}
    \\
    \lag_5^{\mathrm{red}}&=
    \frac{1}{2} \rfive{\phi \Box}{abc} (D_\mu D^\mu \phi_a) \phi_b \phi_c
    +\bigg[ 
    \frac{1}{2} \rfive{\psi}{ij} (D_\mu \psi_i)^T C D^\mu \psi_j
    +\rfive{\psi \phi}{ija} \bar{\psi}_i \mathrm{i} \slashed{D} \psi_j \phi_a
    +\mathrm{h.c.}\bigg ],\label{L5red}
\end{align}
 where we have used the standard definition
\begin{equation}
    \sigma^{\mu\nu}=\frac{\mathrm{i}}{2} [\gamma^\mu,\gamma^\nu].
\end{equation}
The following symmetry properties are satisfied by the Wilson coefficients
\begin{align}
    \afive{\psi F}{Aij}=&-\afive{\psi F}{Aji}, 
    \quad 
\afive{\psi \phi^2}{ijab}=\afive{\psi \phi^2}{jiab}
=\afive{\psi \phi^2}{ijba},
\\
\afive{\phi F}{ABa}=&\afive{\phi F}{BAa}, \quad 
\afive{\phi \widetilde{F}}{ABa}=
\afive{\phi \widetilde{F}}{BAa}, \quad
\afive{\phi}{abcde}=\mbox{fully symmetric},
\\
\rfive{\psi}{ij}=&\rfive{\psi}{ji}, 
\quad \rfive{\phi \Box}{abc}
=\rfive{\phi \Box}{acb}.
\end{align}
Again, all couplings not involving fermions are real.

At mass dimension 6 we have
\begin{align}
\lag_6^{\mathrm{phys}}&=
\frac{1}{3!} \asix{3F}{ABC} (F^A)^{~\nu}_\mu (F^B)^{~\rho}_\nu (F^C)^{~\mu}_\rho
+\frac{1}{3!} \asix{3\widetilde{F}}{ABC} (F^A)^{~\nu}_\mu (F^B)^{~\rho}_\nu (\widetilde{F}^C)^{~\mu}_\rho
\nonumber\\
&+ \frac{1}{4} \asix{\phi F}{ABab} F^{A}_{\mu\nu} F^{B\, \mu\nu} \phi_a \phi_b
+ \frac{1}{4} \asix{\phi\widetilde{F}}{ABab} F^{A}_{\mu\nu} \widetilde{F}^{B\, \mu\nu} \phi_a \phi_b
\nonumber \\
&+ \frac{1}{3} \asix{\phi D}{abcd} \left[(D_{\mu}\phi_{a})(D^{\mu}\phi_{b})\phi_{c}\phi_{d}+\left(ab\leftrightarrow cd\right)-\frac{1}{2}\left(a\leftrightarrow c\right)-\frac{1}{2}\left(a\leftrightarrow d\right)-\frac{1}{2}\left(b\leftrightarrow c\right)-\frac{1}{2}\left(b\leftrightarrow d\right)\right]
\nonumber \\
&+ \frac{1}{6!} \asix{\phi}{abcdef} \phi_a \phi_b \phi_c \phi_d \phi_e \phi_f
+
\frac{1}{2}\asix{\phi \psi}{ijab} \bar{\psi}_i \gamma^\mu \psi_j
[\phi_a D_\mu \phi_b - \phi_b D_\mu \phi_a]
+
\frac{1}{4} \asix{\bar{\psi}\psi}{ijkl}
(\bar{\psi}_i \gamma^\mu \psi_j)
(\bar{\psi}_k \gamma_\mu \psi_l)
\nonumber \\&
+ \bigg[ \frac{1}{2} \asix{\psi F}{Aija} F^A_{\mu\nu} \psi_i^T C \sigma^{\mu\nu} \psi_j \phi_a
+\frac{1}{2!3!} \asix{\psi \phi}{ijabc} \psi_i^T C \psi_j \phi_a \phi_b \phi_c
+ \frac{1}{4!} \asix{\psi \psi}{ijkl} (\psi_i^T C \psi_j) (\psi_k^T C \psi_l) +\mathrm{h.c.}\bigg], \label{L6phys}
\end{align}
\begin{align}
\lag_6^{\textrm{red}} & =
\frac{1}{2!}\rsix{2F}{AB} (D_\mu F^{A\, \mu \nu}) (D^\rho F^B_{\rho \nu})
+\frac{1}{2!}\rsix{FD\phi}{Aab}\left(D_{\nu}F^{A,\mu\nu}\right) \left[\left(D_{\mu}\phi_{a}\right)\phi_{b}-\left(a\leftrightarrow b\right)\right]
\nonumber \\
 & 
+\frac{1}{2!}\rsix{D\phi}{ab}\left(D_{\mu}D^{\mu}\phi_{a}\right)(D_{\nu}D^{\nu}\phi_{b})
 +\frac{1}{3!}\rsix{\phi D}{abcd}\left(D_{\mu}D^{\mu}\phi_{a}\right)\phi_{b}\phi_{c}\phi_{d}
 \nonumber \\
  & 
 +\rsix{D F\psi}{Aij} D^\nu F_{\mu\nu}^{A}
 \overline{\psi_{i}}\gamma^{\mu} \psi_{j} 
  +\rsix{F\psi}{Aij}F_{\mu\nu}^{A}\overline{\psi_{i}}\gamma^{\mu}
\mathrm{i}  \overleftrightarrow{D}^\nu
 \psi_{j} 
 +\rsix{\widetilde{F}\psi}{Aij}\widetilde{F}_{\mu\nu}^{A}\overline{\psi_{i}}\gamma^{\mu}
\mathrm{i}  \overleftrightarrow{D}^\nu
 \psi_{j} 
 \nonumber \\
 &+\frac{1}{2!}
 \rsix{\phi \psi 1}{ijab}
\left(\bar{\psi}_{i}i \overleftrightarrow{\slashed{D}}\psi_{j}\right)\phi_{a}\phi_{b}
+ \frac{1}{2!}\rsix{\phi \psi 2}{ijab} \overline{\psi}_{i}\gamma^{\mu}\psi_{j} D_{\mu}\left(\phi_{a}\phi_{b}\right)
 +\rsix{\psi D}{ij}i\overline{\psi}_{i}
 \{D_{\mu}D^{\mu},\slashed{D} \}\psi_{j}
\nonumber \\
 & +\left\{ \frac{1}{2!}\rsix{\psi\phi D1}{ija}
 \psi_{i}^{T}C\psi_{j}\left(D_{\mu}D^{\mu}\phi_{a}\right)
 +\frac{1}{2!}\rsix{\psi\phi D2}{ija}\left(D^{\mu}\psi_{i}\right)^{T}C i \sigma^{\mu\nu}\left(D_{\nu}\psi_{j}\right)\phi_{a}\right.\nonumber \\
 & +\left.\frac{1}{2!}\rsix{\psi\phi D3}{ija}\left(D_{\mu}D^{\mu}\psi_{i}\right)^{T}C\psi_{j}\phi_{a}+\textrm{h.c.}\right\}.  \label{L6red}
\end{align}

The tensors appearing above have the following symmetries:
\begin{align}
\asix{3F}{ABC} & =\textrm{\textrm{fully anti-symmetric}},\in\mathbb{R},\\
\asix{3\widetilde{F}}{ABC} & =\textrm{\textrm{fully anti-symmetric}},\in\mathbb{R},\\
\asix{\phi\psi}{ijab} & =-\asix{\phi\psi}{ijba}=[\asix{\phi\psi}{jiab}]^\ast,\\
\asix{\bar{\psi}\psi}{ijkl} & =\asix{\bar{\psi}\psi}{kjil}=
\asix{\bar{\psi}\psi}{ilkj}=
[\asix{\bar{\psi}\psi}{jilk}]^\ast,\\
\asix{\phi D}{abcd}&=\asix{\phi D}{bacd}=\asix{\phi D}{abdc}=\asix{\phi D}{cdab}\textrm{ and } \nonumber \\
\asix{\phi D}{abcd}&+\asix{\phi D}{adbc}+\asix{\phi D}{acdb}=0,\asix{\phi D}{abcd}\in\mathbb{R},\\
\asix{\phi F}{ABab} & =\asix{\phi F}{BAab}=
\asix{\phi F}{ABba} \in\mathbb{R},\\
\asix{\phi \widetilde{F}}{ABab} & =\asix{\phi \widetilde{F}}{BAab}=
\asix{\phi \widetilde{F}}{ABba} \in\mathbb{R},\\
\asix{\phi}{abcdef} & =\textrm{\textrm{fully symmetric}}\in\mathbb{R},\\
\asix{\psi\psi}{ijkl}
 & =\asix{\psi\psi}{jikl}
=\asix{\psi\psi}{ijlk}
=\asix{\psi\psi}{klij}
\textrm{ and } \nonumber \\
\asix{\psi\psi}{ijkl} &
+\asix{\psi\psi}{iljk}
+\asix{\psi\psi}{iklj}=0,\\
\asix{\psi F}{Aija} & =-\asix{\psi F}{Ajia},\\
\asix{\psi \phi}{ijabc} & =\textrm{fully symmetric in \ensuremath{\left(i,j\right)} and also \ensuremath{\left(a,b,c\right)}},\\
\rsix{\psi D}{ij} & =[\rsix{\psi D}{ji}]^\ast,\\
\rsix{D\phi}{ab} & =\rsix{D\phi}{ba}\in\mathbb{R},\\
\rsix{2F}{AB} & =\rsix{2F}{BA}\in\mathbb{R},\\
\rsix{DF\psi}{Aij} & =[\rsix{DF\psi}{Aji}]^\ast,\\
\rsix{F\psi}{Aij} & =[\rsix{F\psi}{Aji}]^\ast,\\
\rsix{\widetilde{F}\psi}{Aij} & =[\rsix{\widetilde{F}\psi}{Aji}]^\ast,\\
\rsix{FD\phi}{Aab} & =-\rsix{FD\phi}{Aba}\in\mathbb{R},\\
\rsix{\phi \psi x}{ijab} & 
=\rsix{\phi \psi x}{ijba}=\left[\rsix{\phi\psi x}{jiab}\right]^{*} \textrm{ for }x=1,2, \\
\rsix{\phi D}{abcd} & =\textrm{\textrm{fully symmetric in \ensuremath{\left(b,c,d\right)}}}\in\mathbb{R},\\
\rsix{\psi \phi D1}{ija} & 
=\rsix{\psi \phi D1}{jia},\\
\rsix{\psi \phi D2}{ija} & 
=\rsix{\psi \phi D2}{jia},\\
\rsix{\psi \phi D3}{ija} & =\textrm{no restrictions}.
\end{align}
Note that on top of these constraints, all couplings must respect linear equations involving $t^A$, $\theta^A$ and/or $f^{ABC}$ due to gauge invariance.

In the following, whenever possible we will keep the fermionic indices implicit preserving the correct order in the matrix multiplication. Also the dagger symbol will include the transposition of fermionic indices.

\subsection{Evanescent operators}

The EFT Lagrangian defined by Eqs. (\ref{L1}, \ref{L5phys}, \ref{L5red}, \ref{L6phys}, \ref{L6red}) is enough to parametrize any off-shell Green's function in $d=4$ dimensions, as needed to compute the one-loop renormalization of the EFT itself. However, if we are interested in one-loop finite matching, evanescent structures can become relevant and have to be taken into account. In particular, four-fermion operators related to the ones in our basis via Fierz identities can be generated at tree level and, in their reduction to the physical basis, leave an evanescent shift (see~\cite{Aebischer:2022aze,Fuentes-Martin:2022vvu} for a detailed discussion in the case of the SMEFT at mass dimension 6). The complete set of redundant (in $d=4$) four-fermion operators that will induce the corresponding evanescent shifts reads
\begin{align}
    \mathcal{L}^{(6)}_{\mathrm{ev}}&=\frac{1}{4}(r_{\psi\overline{\psi}}^{(6)})_{ijkl}(\psi_i^TC\psi_j) (\overline{\psi}_k C\overline{\psi}^T_l)+\frac{1}{4}(r_{\psi \overline{\psi}2}^{(6)})_{ijkl}(\psi_i^TC\sigma_{\mu\nu}\psi_j) (\overline{\psi}_k \sigma^{\mu\nu}C\overline{\psi}^T_l)\\
    &+\bigg(\frac{1}{4!}(r_{\psi\psi 2}^{(6)})_{ijkl}(\psi_i^TC\sigma_{\mu\nu}\psi_j) (\psi_k^T C \sigma^{\mu\nu}\psi_l) + \mathrm{h.c.}\bigg),
\end{align}
whose coefficients satisfy the following symmetry properties:
\begin{align}
    (r_{\psi\overline{\psi}}^{(6)})_{ijkl}&=(r_{\psi\overline{\psi}}^{(6)})_{jikl}=(r_{\psi\overline{\psi}}^{(6)})_{ijlk}=(r_{\psi\overline{\psi}}^{(6)})_{klij}^\ast, \\
    (r_{\psi\overline{\psi}2}^{(6)})_{ijkl}&=-(r_{\psi\overline{\psi}2}^{(6)})_{jikl}=-(r_{\psi\overline{\psi}2}^{(6)})_{ijlk}=(r_{\psi\overline{\psi}2}^{(6)})_{klij}^\ast, \\ 
    (r_{\psi\psi2}^{(6)})_{ijkl}&=-(r_{\psi\psi2}^{(6)})_{jikl}=-(r_{\psi\psi2}^{(6)})_{ijlk}=(r_{\psi\psi 2}^{(6)})_{klij}. 
\end{align}
In $d=4$, they can be expressed in terms of operators in the physical basis:
\begin{align}
(\psi_i^TC\psi_j) (\overline{\psi}_k C\overline{\psi}^T_l)&=\frac{1}{2}(\overline{\psi}_l\gamma_{\mu}\psi_i) (\overline{\psi}_k \gamma^{\mu}\psi_j),\\
(\psi_i^TC\sigma_{\mu\nu}\psi_j) (\overline{\psi}_k \sigma^{\mu\nu}C\overline{\psi}^T_l)&=0,\\
(\psi_i^TC\sigma_{\mu\nu}\psi_j) (\psi_k^T C \sigma^{\mu\nu}\psi_l)&=-8(\psi_i^TC\psi_l)(\psi_k^TC\psi_j)-4(\psi_i^TC\psi_j)(\psi_k^TC\psi_l).
\end{align}
so that the physical coefficients experience a shift:
\begin{align}
(a_{\overline{\psi}\psi}^{(6)})_{ijkl}&\rightarrow(a_{\overline{\psi}\psi}^{(6)})_{ijkl}+\frac{1}{2}(r_{\psi\overline{\psi}}^{(6)})_{jlki},\\
(a_{\psi\psi}^{(6)})_{ijkl}&\rightarrow(a_{\psi\psi}^{(6)})_{ijkl}-4(r_{\psi \overline{\psi}2}^{(6)})_{ijkl}-8(r_{\psi\overline{\psi}2}^{(6)})_{ilkj}.
\end{align}
Note that it is after this shift is implemented that the corresponding WCs have the symmetries stated in the previous section.

This is the only relevant shift from these evanescent structures relevant for this work, as we are only interested in the one-loop renormalization of this general EFT. For finite matching or two loop RGE calculations one would need to include extra shifts when the evanescent structures hit a UV pole, see~\cite{Fuentes-Martin:2022vvu} for details.
These will be provided in a companion paper~\cite{companion_fermions} together with the renormalization of the fermionic operators.

\subsection{A note on operators with mixed symmetries}

Out of the operators defined above, there are two that have mixed symmetries under the permutation of indices for identical particles, namely
\begin{align}
(\mathcal{O}^{(6)}_{\phi D})_{abcd} & \equiv \frac{1}{3}  \left[(D_{\mu}\phi_{a})(D^{\mu}\phi_{b})\phi_{c}\phi_{d}+\left(ab\leftrightarrow cd\right)-\frac{1}{2}\left(a\leftrightarrow c\right)-\frac{1}{2}\left(a\leftrightarrow d\right)-\frac{1}{2}\left(b\leftrightarrow c\right)-\frac{1}{2}\left(b\leftrightarrow d\right)\right],
 \\
(\mathcal{O}^{(6)}_{\psi \psi})_{ijkl}& \equiv
\frac{1}{4}
 (\psi_i^T C \psi_j) (\psi_k^T C \psi_l).
\end{align}
We note that the symmetry of both these operators can be described with a single equation 
\begin{equation}
P \mathcal{O}_{abcd}= \mathcal{O}_{abcd},
\end{equation}
where $P$ is the following element of the $S_4$ algebra:
\begin{equation}
P=\sum_{\pi \in S_4} c_\pi \pi,
\end{equation}
with
\begin{align}
c_{1}=&c_{(12)}=c_{(34)}=c_{(1324)}=c_{(1423)}=c_{(12)(34)}=c_{(13)(24)}=c_{(14)(23)}=\frac{1}{12}, \\
c_{(13)}=&c_{(14)}=c_{(23)}=c_{(24)}=c_{(123)}=c_{(124)}=c_{(132)}=c_{(134)}
\nonumber \\
=&c_{(142)}=c_{(143)}=c_{(234)}=c_{(243)}=c_{(1234)}=c_{(1243)}=c_{(1342)}=c_{(1432)}=-\frac{1}{24}\,.
\end{align}
This $P$ is a projector, i.e. $P^2=P$, and furthermore it is self-adjoint in the sense that
\begin{equation}
P^\prime \equiv \sum_{\pi \in S_4} c_\pi (\pi\RF{^{-1}})
= P.
\end{equation}
The last property is important because it implies that the Wilson coefficients $\alpha_{abcd}$ contracted with the operators $(\mathcal{O}^{(6)}_{\phi D})_{abcd}$ and $(\mathcal{O}^{(6)}_{\psi \psi})_{abcd}$ can be described with the same projector $P$ as the operators themselves,

\begin{equation}
P \alpha_{abcd}= \alpha_{abcd}\,.\label{eq:P-proj}
\end{equation}

This single symmetry constraint on the Wilson coefficients is equivalent to the set of equations given in Section \ref{sec_general_EFT}. Importantly, we can use the fact that --- aside from gauge invariance --- eq. \eqref{eq:P-proj} is the only constraint on $\alpha$ to easily build the most general tensor that obeys it. To do so, for each of the two operators one can start with a completely general rank-4 tensor with no symmetries $\tilde{\alpha}_{abcd}$ and simply project out with $P$ the unphysical parts:
\begin{equation}
\alpha_{mnop} \equiv P \tilde{\alpha}_{mnop}\,.
\end{equation}
With this procedure, the Wilson coefficients of the two operators satisfy the correct symmetries.

It should be noted that if the tensor $\tilde{\alpha}_{mnop}$ has certain symmetries, the action of the projector can simplify significantly. In particular, if it is symmetric under $m\leftrightarrow n$ and under $o\leftrightarrow p$ then we have
\begin{equation}
    P\tilde{\alpha}_{mnop}= \frac{1}{3} \left[ 
    \tilde{\alpha}_{mnop} + (mn \leftrightarrow op) 
    - \frac{1}{2} \Big( 
    (m \leftrightarrow o)
    +(m \leftrightarrow p)
    +(n \leftrightarrow o)
    +(n \leftrightarrow p)\Big)\right].
\end{equation}
This is the form we have used in this article.

\subsection{Dealing with semi-simple gauge groups and $U(1)$ mixing}

In presenting our results we depart with the well established convention
of introducing new names for combinations of couplings \cite{Machacek:1983tz,Machacek:1983fi,Machacek:1984zw,Martin:1993zk,Yamada:1994id,Schienbein:2018fsw,Luo:2002ti}.
In particular, in all cases we write down explicitly the representation
matrices $\left(R^{A}\right)_{ab}=\theta_{ab}^{A}$ (for scalars),
$t_{ab}^{A}$ (for fermions) or $-if^{Aab}$ (for vectors). This has
the advantage of making it straightforward to apply our results to
models with a semi-simple group: the gauge coupling constant $g$
which always accompanies the matrices $R^{A}$ can depend on the index
$A$, hence one must make the replacement
\begin{equation}
g\theta^{A},g t^{A},g f^{ABC}\rightarrow g_{A}\theta^{A},g_{A}t^{A},g_{A}f^{ABC}
\end{equation}
in all formulas.

Gauge invariance forces the $a_{KF}$ matrix to be diagonal, unless
there are several $U(1)$'s. In the latter case, there might be $U(1)$
mixing \cite{Holdom:1985ag} and one can tackle it by promoting $g$ to a matrix rather
than a list of couplings \cite{delAguila:1988jz, delAguila:1995rb, Fonseca:2013jra}:
\begin{equation}
g_{B}R^{B}V_{\mu}^{B}\rightarrow g_{AB}R^{A}V_{\mu}^{B},\quad R=\theta,t,f\,.
\end{equation}
In our expressions, one should change 
\begin{equation}
g \theta^{A},g t^{A},g f^{ABC}\rightarrow g_{XA}\theta^{X},g_{XA}t^{X},g_{XA}f^{XBC}\,.
\end{equation}

\section{Reduction to the physical basis\label{sec_reduction}}

The Green's basis can be reduced to the physical one via field redefinitions. These can be equivalent to the use of equations of motion at the linear order, which is enough for the calculation of the one-loop beta functions (non-linear terms in the redundant operators are higher than one-loop order). Nevertheless, we reproduce here the exact reduction up to mass dimension 6, including terms that are quadratic in mass dimension 5 operators, for general use.
The following field redefinitions can be used, after canonical normalization, to eliminate the redundant operators in favor of the physical ones
\begin{align}
\phi_a &\to \phi_a + \frac{1}{2}(r_{\phi\Box}^{(5)})_{abc}\phi_b\phi_c,\\
\psi &\to \psi 
-\frac{i}{2} \rfive{\psi}{}^\dagger \slashed{D}\psi^c
+\frac{i}{8} \rfive{\psi}{}^\dagger m_\psi \rfive{\psi}{}^\dagger  \slashed{D}\psi^c.
\end{align}
These field redefinitions modify the kinetic term for the fermions, which have to be canonically normalized again. After that, the following field redefinition eliminates all the remaining dimension 5 redundant operators,
\begin{equation}
\psi \to \psi 
- \rfive{\psi\phi}{a}^\dagger\psi\phi_a,
\end{equation}
where we have left the fermionic indices implicit.

At mass dimension 6, once the mass dimension 5 redundant operators have been reduced~\footnote{This reduction includes, at quadratic order, contributions to operators of mass dimension 6, that have to be reduced at this step.}, we can use directly the equations of motion from $\mathcal{L}_{d\leq 4}$, that read,
\begin{align}
    \mathrm{i} \slashed{D} \psi =&\Big[ (m_\psi)^\dagger
    + Y^\dagger_{a} \phi_a 
    \Big] C \bar{\psi}^T,
    \\
    D^2\phi_a =&\, \eta_a-(m_\phi^2)_{ab} \phi_b
    -\frac{1}{2} \Big[ \psi^T Y_a C \psi  + 
     \bar{\psi} Y_a^\dagger C \bar{\psi}^T
    \Big]
    -\frac{\kappa_{abc}}{2} \phi_b \phi_c
    -\frac{\lambda_{abcd}}{3!} \phi_b \phi_c \phi_d, \\
    (D^\nu F_{\nu\mu})^A =&
-\mathrm{i} g_A \theta^A_{ab} \phi_a D_\mu \phi_b
    - g_A  \bar{\psi} t^A \gamma_{\mu} \psi.
\end{align}

The application of these field redefinitions and equations of motion results in the following shifts for the physical WCs,
\begin{align}
(m_{\phi}^2)_{ab}&\to 
(m_{\phi}^2)_{a b} 
+ \eta_{c}\,\rfive{\phi\Box}{cab}
- (m_{\phi}^2)_{a c}\,\rsix{D\phi}{c d}\,(m_{\phi}^2)_{db} 
- \kappa_{a b c}\,\rsix{D\phi}{c d}\,\eta_{d},\\
\eta_a& \to 
\eta_{a} 
- (m_{\phi}^2)_{a b}\,\rsix{D\phi}{bc}\,\eta_{c},\\
(m_\psi)_{ij}&\to(m_\psi)_{ij}+\frac{1}{2}\Big\{
    -\Big[m_\psi \rfive{\psi}{}^\dagger m_\psi\Big]_{ij}
    +\Big[m_\psi \rfive{\psi}{}^\dagger m_\psi \rfive{\psi}{}^\dagger m_\psi \Big]_{ij}
    \nonumber\\
    &\phantom{\to(m_\psi)_{ij}+\frac{1}{2}\Big\{}
     -\frac{7}{2}\Big[m_\psi m_\psi^\dagger \rfive{\psi}{} \rfive{\psi}{}^\dagger m_\psi \Big]_{ij}
     +\eta_a \rsix{\psi\phi D1}{ija}+\frac{1}{2}\eta_a \rsix{\psi\phi D2}{ija}\nonumber\\
    &\phantom{\to(m_\psi)_{ij}+\frac{1}{2}\Big\{}
    -\eta_a Y_{ijb}\rsix{D\phi}{ba}
    -2\Big[m_\psi \rsix{\psi D}{} m_\psi^\dagger m_\psi \Big]_{ij} + (i\leftrightarrow j)\Big\},\\
\kappa_{abc}&\to \kappa_{abc} 
+ \frac{1}{3!}\sum_{\mathrm{perm}}\Bigg\{ 
3\,(m_{\phi}^2)_{a d}\,\rfive{\phi\Box}{d b c} 
+ 3 \rfive{\phi\Box}{d c e}\,\rfive{\phi\Box}{e a b}\,\eta_{d} 
+ \frac{17}{12}\,\rfive{\phi\Box}{e ad}\,\rfive{\phi\Box}{e b c}\,\eta_{d}
\nonumber\\ & 
\phantom{\kappa_{abc} + \frac{1}{6}\sum_{\mathrm{perm}}\Bigg\{ }
+\rsix{\phi D}{d a b c}\,\eta_{d}
- 3\,(m_{\phi}^2)_{a d}\,\rsix{D\phi}{de}\,\kappa_{b c e} 
- \rsix{D\phi}{de}\,\eta_{e}\,\lambda_{a b c d}
\Bigg\},\\
\lambda_{abcd}& \to \lambda_{abcd}
+\frac{1}{4!}\sum_{\mathrm{perm}}\Bigg\{ 
6\,\rfive{\phi\Box}{e a d}\,\kappa_{b c e}
+ \frac{3}{2}\,(m_{\phi}^2)_{f e}\,\rfive{\phi\Box}{f a b}\,\rfive{\phi\Box}{e c d}
\nonumber \\ &
\phantom{\lambda_{abcd}+\frac{1}{24}\sum_{\mathrm{perm}}\Bigg\{} 
+ 12\,(m_{\phi}^2)_{a f}\,\rfive{\phi\Box}{f b e}\,\rfive{\phi\Box}{e c d} 
+ \frac{17}{3}\,(m_{\phi}^2)_{a f}\,\rfive{\phi\Box}{e b f}\,\rfive{\phi\Box}{e c d} 
\nonumber \\&
\phantom{\lambda_{abcd}+\frac{1}{24}\sum_{\mathrm{perm}}\Bigg\{}
- 3\,\rsix{D\phi}{f e}\,\kappa_{a b f}\,\kappa_{c d e} 
- 4\,(m_{\phi}^2)_{a f}\,\rsix{D\phi}{f e}\,\lambda_{b c d e}
+ 4\,(m_{\phi}^2)_{a e}\,\rsix{\phi D}{e b c d} 
\Bigg\},\\
Y_{ija}&\to Y_{ija}+\frac{1}{2}\Bigg\{-2\Big[m_\psi \rfive{\psi\phi}{a}^\dagger\Big]_{ij}-2\Big[m_\psi \rfive{\psi}{}^\dagger Y_a\Big]_{ij}+2\Big[ m_\psi m_\psi^\dagger \rfive{\psi}{}\rfive{\psi\phi}{a}^\dagger\Big]_{ij}\nonumber\\
&\phantom{\to Y_{ij}^a+\frac{1}{2}\Bigg\{}+2\Big[ m_\psi \rfive{\psi}{}^\dagger m_\psi \rfive{\psi\phi}{a}^\dagger\Big]_{ij}+2\Big[ m_\psi \rfive{\psi\phi}{a}^\dagger \rfive{\psi}{}^\dagger m_\psi\Big]_{ij} + \Big[ m_\psi \rfive{\psi}{}^\dagger Y_a \rfive{\psi}{}^\dagger m_\psi\Big]_{ij}\nonumber\\
    &\phantom{\to Y_{ij}^a+\frac{1}{2}\Bigg\{}-\frac{15}{8}\Big[ m_\psi m_\psi^\dagger \rfive{\psi}{} \rfive{\psi}{}^\dagger Y_a \Big]_{ij} + \frac{15}{4}\Big[ m_\psi m_\psi^\dagger Y_a \rfive{\psi}{}^\dagger \rfive{\psi}{}\Big]_{ij}\nonumber\\
    &\phantom{\to Y_{ij}^a+\frac{1}{2}\Bigg\{}+\frac{15}{2}\Big[m_\psi \rfive{\psi}{}^\dagger m_\psi \rfive{\psi}{}^\dagger Y_a \Big]_{ij}-\frac{13}{8}\Big[ m_\psi \rfive{\psi}{}^\dagger \rfive{\psi}{} m_\psi^\dagger Y_a \Big]_{ij}\nonumber\\
    &\phantom{\to Y_{ij}^a+\frac{1}{2}\Bigg\{}+(m_\phi^2)_{ab} \rsix{\psi\phi D1}{ijb} + \frac{1}{2}(m_\phi^2)_{ab} \rsix{\psi\phi D2}{ijb} - (m_\phi^2)_{ab} Y_{ij}^c \rsix{D\phi}{bc}\nonumber\\
    &\phantom{\to Y_{ij}^a+\frac{1}{2}\Bigg\{}+2 \Big[m_\psi m_\psi^\dagger Y_a \rsix{\psi D}{}\Big]_{ij} - \Big[m_\psi \rsix{\psi D}{} m_\psi^\dagger Y_a\Big]_{ij} - \Big[Y_a \rsix{\psi D}{} m_\psi^\dagger m_\psi \Big]_{ij}\nonumber\\
    &\phantom{\to Y_{ij}^a+\frac{1}{2}\Bigg\{}-\Big[m_\psi m_\psi^\dagger \rsix{\psi\phi D2}{a}\Big]_{ij} + \Big[m_\psi \rsix{\psi\phi D2}{a}^\dagger m_\psi \Big]_{ij} +\Big[m_\psi m_\psi^\dagger \rsix{\psi\phi D3}{a}\Big]_{ij}\nonumber\\
    &\phantom{\to Y_{ij}^a+\frac{1}{2}\Bigg\{}+(i \leftrightarrow j)\Bigg\},\\
\afive{\phi}{abcde}& \to \afive{\phi}{abcde}
+\frac{1}{5!}\sum_{\mathrm{perm}} \Bigg\{
- 10\,\rfive{\phi\Box}{f a e}\,\lambda_{b c d f}
- 30\,\rfive{\phi\Box}{f c d}\,\rfive{\phi\Box}{g b f}\,\kappa_{a e g} 
\nonumber \\ & 
\phantom{\to \afive{\phi}{abcde} +\frac{1}{120}\sum_{\mathrm{perm}} \Bigg\{}
- 15\,\rfive{\phi\Box}{f a b}\,\rfive{\phi\Box}{g d e}\,\kappa_{c f g}
- \frac{85}{6}\,\rfive{\phi\Box}{f b g}\,\rfive{\phi\Box}{f c d}\,\kappa_{a e g} \nonumber \\ &
\phantom{\to \afive{\phi}{abcde} +\frac{1}{120}\sum_{\mathrm{perm}} \Bigg\{}
- 10\,\rsix{\phi D}{fb c d}\,\kappa_{afe}  
+ 10\,\rsix{D\phi}{fg}\,\kappa_{abf}\,\lambda_{cdeg} 
\Bigg\},\\
\afive{\psi F}{ijA}&\to\afive{\psi F}{ijA}+ \frac{1}{2}
\Bigg\{
- \frac{1}{2}g\Big[\rfive{\psi}{}  t^{A}\Big]_{i j}- 2\Big[\,m_{\psi}  \rfive{\psi}{}^{\dagger}  \afive{\psi F}{A}\Big]_{i j}  
\nonumber\\ &
\phantom{\to\afive{\psi F}{ijA}+ \frac{1}{2} \Bigg\{}
+ \frac{5}{8}\,g\Big[m_{\psi} \rfive{\psi}{}^{\dagger}  \rfive{\psi}{}  t^{A}\Big]_{i j} 
+ \frac{1}{8}g\Big[\rfive{\psi}{}  m_{\psi}^\dagger  \rfive{\psi}{}  t^{A}\Big]_{i j}
\nonumber\\ 
&\phantom{\to\afive{\psi F}{ijA}+ \frac{1}{2} \Bigg\{}- \Big[m_{\psi} \, \rsix{\tilde{F}\psi}{A}\Big]_{i j} 
+ \mathrm{i}\,\Big[m_{\psi}  \rsix{F\psi}{A}\Big]_{i, j} 
-(i \leftrightarrow j)\Bigg\}, 
\\
    \afive{\psi\phi2}{ijab}&\to\afive{\psi\phi2}{ijab}+\frac{1}{4}\Bigg\{- \rfive{\phi\Box}{c a b}\,Y_{i j c} +4 \,\Big[Y_{a}  \rfive{\psi\phi}{b}^{\dagger}\Big]_{i j}+2 \Big[Y_{a} \rfive{\psi}{}^{\dagger} Y_{b} \Big]_{i j} \nonumber\\
    &\phantom{\to\afive{\psi\phi2}{ijab}+\frac{1}{4}\Bigg\{}+ 2\,\rfive{\phi\Box}{c a b}\,\Big[m_{\psi}  \rfive{\psi\phi}{c}^{\dagger}\Big]_{i j} -2\Big[m_{\psi}  \rfive{\psi}{}^{\dagger}  \afive{\psi\phi2}{a b}\Big]_{i j}\nonumber\\
    &\phantom{\to\afive{\psi\phi2}{ijab}+\frac{1}{4}\Bigg\{}+2\rfive{\phi\Box}{c a b}\,\Big[m_{\psi}  \rfive{\psi}{}^{\dagger}  Y_{c}\Big]_{i j}
    &\phantom{\to\afive{\psi\phi2}{ijab}+\frac{1}{4}\Bigg\{}- 2\,\Big[m_{\psi}  \rfive{\psi\phi}{a}  \rfive{\psi\phi}{b}^{\dagger}\Big]_{i j} - 4\,\Big[m_{\psi}  \rfive{\psi\phi}{a}^{\dagger}  \rfive{\psi\phi}{b}^{\dagger}\Big]_{i j} \nonumber\\
    &\phantom{\to\afive{\psi\phi2}{ijab}+\frac{1}{4}\Bigg\{}-\frac{15}{4}\Big[ m_\psi \rfive{\psi}{}^\dagger\rfive{\psi}{}Y^{\dagger}_a Y_b\Big]_{ij}-\frac{1}{2}\Big[ m_\psi Y^{\dagger}_a\rfive{\psi}{}\rfive{\psi}{}^\dagger Y_b \Big]_{ij}\nonumber\\
    &\phantom{\to\afive{\psi\phi2}{ijab}+\frac{1}{4}\Bigg\{}-\frac{15}{4}\Big[ m_\psi Y^{\dagger}_a Y_b\rfive{\psi}{}^\dagger \rfive{\psi}{} \Big]_{ij}-\frac{15}{2}\Big[ Y_a m_\psi^\dagger Y_b \rfive{\psi}{}^\dagger \rfive{\psi}{} \Big]_{ij}\nonumber\\
    &\phantom{\to\afive{\psi\phi2}{ijab}+\frac{1}{4}\Bigg\{}-4\Big[m_\psi \rfive{\psi}{}^\dagger Y_a \rfive{\psi}{}^\dagger Y_b \Big]_{ij}-3\Big[ Y_a \rfive{\psi}{}^\dagger m_\psi \rfive{\psi}{}^\dagger Y_b \Big]_{ij}\nonumber\\
    &\phantom{\to\afive{\psi\phi2}{ijab}+\frac{1}{4}\Bigg\{}-4\Big[ m_\psi \rfive{\psi}{}^\dagger Y_a \rfive{\psi\phi}{b}^\dagger \Big]_{ij}-4\Big[ m_\psi \rfive{\psi\phi}{a}^\dagger \rfive{\psi}{}^\dagger Y_b  \Big]_{ij}\nonumber\\
    &\phantom{\to\afive{\psi\phi2}{ijab}+\frac{1}{4}\Bigg\{}-4\Big[ Y_a  \rfive{\psi}{}^\dagger m_\psi \rfive{\psi\phi}{b}^\dagger \Big]_{ij} -4\Big[ Y_a  \rfive{\psi\phi}{b}^\dagger \rfive{\psi}{}^\dagger m_\psi  \Big]_{ij}\nonumber\\
    &\phantom{\to\afive{\psi\phi2}{ijab}+\frac{1}{4}\Bigg\{}-2\Big[ m_\psi \rfive{\psi\phi}{a}\rfive{\psi}{}^\dagger Y_b \Big]_{ij}\nonumber\\
    &\phantom{\to\afive{\psi\phi2}{ijab}+\frac{1}{4}\Bigg\{}-2\Big[ m_\psi Y^{\dagger}_a \rfive{\psi}{} \rfive{\psi\phi}{b}^\dagger \Big]_{ij} -4\Big[  Y_a m_\psi^\dagger \rfive{\psi}{} \rfive{\psi\phi}{b}^\dagger \Big]_{ij}\nonumber\\
    &\phantom{\to\afive{\psi\phi2}{ijab}+\frac{1}{4}\Bigg\{}-2\Big[m_\psi \rfive{\psi\phi}{a}\rfive{\psi\phi}{b}^\dagger \Big]_{ij}-4\Big[ m_\psi \rfive{\psi\phi}{a}^\dagger \rfive{\psi\phi}{b}^\dagger \Big]_{ij}\nonumber\\
    &\phantom{\to\afive{\psi\phi2}{ijab}+\frac{1}{4}\Bigg\{}-2\Big[ \overline{\rfive{\psi\phi}{a}} m_\psi  \rfive{\psi\phi}{b}^\dagger \Big]_{ij}\nonumber\\
    &\phantom{\to\afive{\psi\phi2}{ijab}+\frac{1}{4}\Bigg\{}+ \rsix{\psi\phi D1}{i j c}\,\kappa_{a b c} + \frac{1}{2}\rsix{\psi\phi D2}{i j c}\,\kappa_{a b c}+ \rsix{D\phi}{c d}\,Y_{i j}^{c}\,\kappa_{a b d}\nonumber\\
    &\phantom{\to\afive{\psi\phi2}{ijab}+\frac{1}{4}\Bigg\{}+ 2\,\Big[m_{\psi}  \rsix{\phi\psi1}{a b}\Big]_{i j} - 2\,\mathrm{i}\,\Big[m_{\psi}  \rsix{\phi\psi2}{a b}\Big]_{i j}- \Big[m_{\psi}  \rsix{\psi\phi D2}{a}^\dagger  Y_{b}\Big]_{i j} \nonumber\\
    &\phantom{\to\afive{\psi\phi2}{ijab}+\frac{1}{4}\Bigg\{}+ \Big[m_{\psi}  Y^{\dagger}_b  \rsix{\psi\phi D2}{a}\Big]_{i j}+ 2\Big[Y_a m_{\psi}^\dagger    \rsix{\psi\phi D2}{b}\Big]_{i j}\nonumber\\
    &\phantom{\to\afive{\psi\phi2}{ijab}+\frac{1}{4}\Bigg\{}- 2\Big[Y_a \rsix{\psi\phi D2}{b}^\dagger m_{\psi}\Big]_{i j}- \Big[m_{\psi}  \rsix{\psi\phi D3}{a}^\dagger  Y_{b}\Big]_{i j}\nonumber\\
    &\phantom{\to\afive{\psi\phi2}{ijab}+\frac{1}{4}\Bigg\{} - \Big[m_{\psi}  Y^{\dagger}_b  \rsix{\psi\phi D3}{a}\Big]_{i j} -2\Big[Y_a m_{\psi}^\dagger    \rsix{\psi\phi D3}{b}\Big]_{i j}\nonumber\\
    &\phantom{\to\afive{\psi\phi2}{ijab}+\frac{1}{4}\Bigg\{}-2\Big[m_\psi \rsix{\psi D}{} Y^{\dagger}_a Y_b \Big]_{ij}-2\Big[m_\psi Y^{\dagger}_a Y_b \rsix{\psi D}{}  \Big]_{ij}-4\Big[ Y_{a}m_\psi^\dagger Y_a \rsix{\psi D}{}  \Big]_{ij}\nonumber\\
    &\phantom{\to\afive{\psi\phi^2}{ijab}+\frac{1}{4}\Bigg\{}
+(i \leftrightarrow j)+(a \leftrightarrow b)+(i \leftrightarrow j)(a \leftrightarrow b)\Bigg\},
    &\\
    \asix{\phi F}{ABab}&\to \asix{\phi F}{ABab} + \afive{\phi F}{A B c}\,\rfive{\phi\Box}{c a b}\\
    \asix{\phi D}{abcd}&\to \asix{\phi D}{abcd} +\frac{1}{12}\Bigg\{ \rfive{\phi\Box}{e a d}\,\rfive{\phi\Box}{e b c}
    +\rfive{\phi\Box}{e a c}\,\rfive{\phi\Box}{e b d}
    -\rfive{\phi\Box}{e a b}\,\rfive{\phi\Box}{e c d}
    \nonumber\\
    &\phantom{\to \asix{\phi D}{abcd} +\frac{1}{12}\Bigg\{}
    +\mathrm{i}g\theta^A_{ca}\rsix{FD\phi}{Abd} 
    +\mathrm{i}g\theta^A_{cb}\rsix{FD\phi}{Aad}
    +\mathrm{i}g\theta^A_{da}\rsix{FD\phi}{Abc}
    +\mathrm{i}g\theta^A_{db}\rsix{FD\phi}{Aac} 
    \nonumber\\
    &\phantom{\to \asix{\phi D}{abcd} +\frac{1}{12}\Bigg\{}
    -g^2\theta^A_{db}\theta^B_{ca}\rsix{2F}{AB}
    -g^2\theta^A_{da}\theta^B_{cb}\rsix{2F}{AB}\nonumber\\
    &\phantom{\to \asix{\phi D}{abcd} +\frac{1}{12}\Bigg\{}
    +\left(ab\leftrightarrow cd\right)
    -\frac{1}{2}\left(a\leftrightarrow c\right)
    -\frac{1}{2}\left(a\leftrightarrow d\right)
    -\frac{1}{2}\left(b\leftrightarrow c\right)
    -\frac{1}{2}\left(b\leftrightarrow d\right)\Bigg\},\\
    \asix{\phi}{a bc d e f}&\to \asix{\phi}{a bc d e f} +\frac{1}{6!}\sum_{\mathrm{perm}}\Bigg\{ 
    15\,\afive{\phi}{g b c d e}\,\rfive{\phi\Box}{g a f} 
    - \frac{225}{2}\,\rfive{\phi\Box}{g a b}\,\rfive{\phi\Box}{h e f}\,\lambda_{c d g h}
    \nonumber\\ &
    \phantom{\to \asix{\phi}{a bc d e f} +\frac{1}{720}\sum_{\mathrm{perm}}\Bigg\{}- \frac{85}{3}\,\rfive{\phi\Box}{g b h}\, \rfive{\phi\Box}{g c d}\,\lambda_{a e h f} 
    - 60\,\rfive{\phi\Box}{g c d}\,\rfive{\phi\Box}{h b g}\,\lambda_{a e h f}\nonumber \\ &
    \phantom{\to \asix{\phi}{a bc d e f} +\frac{1}{720}\sum_{\mathrm{perm}}\Bigg\{}
    - 20\,\rsix{\phi D}{g b c d}\,\lambda_{a g e f}
    + 10\,\rsix{D\phi}{g h}\,\lambda_{a b c g}\,\lambda_{d e h f}
    \Bigg\},\\
    \asix{\phi\psi}{i ja b}&\to \asix{\phi\psi}{i ja b} +\frac{1}{2}\Bigg\{\mathrm{i}\,\Big[\rfive{\psi\phi}{b}   \rfive{\psi\phi}{a}^\dagger\Big]_{ij} + \frac{\mathrm{i}}{4}\,\Big[Y^{\dagger}_b   \rfive{\psi}{}   \rfive{\psi}{}^\dagger   Y_{a}\Big]_{ij} \nonumber\\
    &\phantom{\to \asix{\phi\psi}{i ja b} +\frac{1}{2}\Bigg\{}+\Bigg[ \frac{15\mathrm{i}}{8}\,\Big[Y^{^\dagger}_b Y_a \rfive{\psi}{}^\dagger \rfive{\psi}{}\Big]_{ij}   \nonumber\\
    &\phantom{\to \asix{\phi\psi}{i ja b} +\frac{1}{2}\Bigg\{}+ \frac{\mathrm{i}}{16}g^2\,\theta^{A}_{a b}\,\Big[t^{A}\rfive{\psi}{}^\dagger   r_{\psi}^{(5)}   \Big]_{ij}- \mathrm{i}\,\Big[\rfive{\psi\phi}{a}   \rfive{\psi}{}^\dagger   Y_{b}\Big]_{ij}\nonumber\\
    &\phantom{\to \asix{\phi\psi}{i ja b} +\frac{1}{2}\Bigg\{}- \mathrm{i}\,\Big[\rsix{\psi D}{}   Y^{\dagger}_a   Y_{b}\Big]_{ij}  
    +\frac{\mathrm{i}}{2}\,\Big[\rsix{\psi\phi D2}{a}^\dagger   Y_{b}\Big]_{ij} -\frac{\mathrm{i}}{2}\,\Big[\rsix{\psi\phi D3}{a}^\dagger   Y_{b}\Big]_{ij}+\mathrm{h.c.}\Bigg]\nonumber\\
    &\phantom{\to \asix{\phi\psi}{i ja b} +\frac{1}{2}\Bigg\{}+ \mathrm{i}g^2\,\rsix{2F}{A B}\,\theta^{A}_{a b}\,t^{B}_{i j} + \mathrm{i}g\,\rsix{DF\psi}{A i j}\,\theta^{A}_{a b} + ig\,\rsix{\tilde{F}\psi}{A i j}\,\theta^{A}_{a b}  - (a\leftrightarrow b)\Bigg\},\\
    \asix{\overline{\psi}\psi}{i j k l}&\to\asix{\overline{\psi}\psi}{i j k l} + \frac{1}{2}\rsix{\psi\psi}{j l i k} + 4\,\rsix{DF\psi}{A i j}\,t^{A}_{k l} + 4\,\rsix{\tilde{F}\psi}{A i j}\,t^{A}_{k l} \nonumber\\
    &+ \rsix{2F}{A B}\,t^{A}_{i j}\,t^{B}_{k l} +\frac{1}{2} \rsix{D\phi}{a b}\,Y_{j la}\,Y^{\dagger}_{i k b} \nonumber\\
    &- \rsix{\psi\phi D1}{i k a}^\dagger\,Y_{j l a} -\rsix{\psi\phi D1}{j l a}\,Y^{\dagger}_{i ka}  \nonumber\\
    &- \frac{1}{2}\rsix{\psi\phi D2}{i k a}^\dagger\,Y_{j la} - \frac{1}{2}\rsix{\psi\phi D2}{j l a}\,Y^{\dagger}_{i k a}, \\
    \asix{\psi F}{A i j a} &\to\asix{\psi F}{A i j a} +\frac{1}{2}\Bigg\{- 2\Big[\afive{\psi F}{A}  \rfive{\psi\phi}{a}^{\dagger}\Big]_{ij}  - 2    \Big[\afive{\psi F}{A} \rfive{\psi}{}^{\dagger}  Y_{a}\Big]_{ij}  \nonumber\\
    &\phantom{\to\asix{\psi F}{A i j a} +\frac{1}{2}\Bigg\{}+ g \Big[\rfive{\psi\phi}{a}^{\dagger T}  \rfive{\psi}{}  t^{A}\Big]_{ij}+\frac{3}{2}g\Big[Y_a t^A \rfive{\psi}{}^\dagger \rfive{\psi}{} \Big]_{ij}\nonumber\\
    &\phantom{\to\asix{\psi F}{A i j a} +\frac{1}{2}\Bigg\{}- \Big[Y_{a}  \rsix{\tilde{F}\psi}{A}\Big]_{ij} + i\,\Big[Y_{a}  \rsix{F\psi}{A}\Big]_{ij} \nonumber\\
    &\phantom{\to\asix{\psi F}{A i j a} +\frac{1}{2}\Bigg\{}+ \frac{1}{2}\Big[\rsix{\psi\phi D3}{a}^T  t^{A}\Big]_{ij} + g\Big[\rsix{\psi D}{}^T  Y_{a}  t^{A}\Big]_{ij}\nonumber\\
    &\phantom{\to\asix{\psi F}{A i j a} +\frac{1}{2}\Bigg\{}-\frac{1}{2}g \Big[\rsix{\psi\phi D2}{a}  t^{A}\Big]_{ij}- (i\leftrightarrow j)\Bigg\}, \\
    \asix{\psi\phi}{i j a b c}&\to\asix{\psi\phi}{i j a b c} +\frac{1}{2!3!}\sum_{\pi(ij),\pi(abc)}\Bigg\{ 3\,\afive{\psi\phi2}{i j d b}\,\rfive{\phi\Box}{d a c} + 3\rfive{\phi\Box}{d c e}\,\rfive{\phi\Box}{e a b}\,Y_{i j}^{d} \nonumber\\
    &\phantom{\to\asix{\psi\phi}{i j a b c} +\frac{1}{2!3!}\sum_{\mathrm{perms}}\Bigg\{}+ \frac{17}{12}\,\rfive{\phi\Box}{e a d}\,\rfive{\phi\Box}{e b c}\,Y_{i j}^{d}- 3\,\Big[\afive{\psi\phi2}{a b}  \rfive{\psi\phi}{c}^\dagger\Big]_{i j} \nonumber\\
    &\phantom{\to\asix{\psi\phi}{i j a b c} +\frac{1}{2!3!}\sum_{\mathrm{perms}}\Bigg\{}+ 6\,\rfive{\phi\Box}{d a b}\,\Big[Y_{c}  \rfive{\psi\phi}{d}^{\dagger}\Big]_{ij} + 6\,\rfive{\phi\Box}{d a b}\,\Big[Y_{d}  \rfive{\psi\phi}{c}^{\dagger}\Big]_{ij} \nonumber\\
    &\phantom{\to\asix{\psi\phi}{i j a b c} +\frac{1}{2!3!}\sum_{\mathrm{perms}}\Bigg\{}- 3\,\Big[  \afive{\psi\phi2}{a b} \rfive{\psi\phi}{c}^{\dagger}\Big]_{ij} - 6\,\Big[Y_{c}  \rfive{\psi}{}^{\dagger}  \afive{\psi\phi2}{a b}\Big]_{ij} \nonumber\\
    &\phantom{\to\asix{\psi\phi}{i j a b c} +\frac{1}{2!3!}\sum_{\mathrm{perms}}\Bigg\{}+ 6\,\rfive{\phi\Box}{d a b}\,\Big[Y_{c}  \rfive{\psi}{}^{\dagger}  Y_{d}\Big]_{ij} \nonumber\\
    &\phantom{\to\asix{\psi\phi}{i j a b c} +\frac{1}{2!3!}\sum_{\mathrm{perms}}\Bigg\{}- 6\,\Big[Y_{c}  \rfive{\psi\phi}{a}  \rfive{\psi\phi}{b}^\dagger\Big]_{ij} - 12\,\Big[Y_{c}  \rfive{\psi\phi}{a}^\dagger  \rfive{\psi\phi}{b}^\dagger\Big]_{ij} \nonumber\\
    &\phantom{\to\asix{\psi\phi}{i j a b c} +\frac{1}{2!3!}\sum_{\mathrm{perms}}\Bigg\{}- 6\,\Big[\rfive{\psi\phi}{a}^{\dagger\,T}  Y_{b}  \rfive{\psi\phi}{c}^\dagger\Big]_{ij} - 12\,\Big[Y_{c}  \rfive{\psi}{}^{\dagger}  Y_{a}  \rfive{\psi\phi}{b}^\dagger\Big]_{ij} \nonumber\\
    &\phantom{\to\asix{\psi\phi}{i j a b c} +\frac{1}{2!3!}\sum_{\mathrm{perms}}\Bigg\{}- 12\,\Big[Y_{c}  \rfive{\psi}{}^{\dagger} \rfive{\psi\phi}{a}^{\dagger\,T}  Y_{b}\Big]_{ij}- 6\,\Big[Y_{c}  \rfive{\psi\phi}{a}  \rfive{\psi}{}^{\dagger}  Y_{b}\Big]_{ij}\nonumber\\
    &\phantom{\to\asix{\psi\phi}{i j a b c} +\frac{1}{2!3!}\sum_{\mathrm{perms}}\Bigg\{}- 6\,\Big[Y_{c}  Y^{\dagger}_b  \rfive{\psi}{}  \rfive{\psi\phi}{a}^\dagger\Big]_{ij} -\frac{45}{4} \,\Big[Y_{c}  Y^{\dagger}_b  Y_{a}  \rfive{\psi}{}^{\dagger}  \rfive{\psi}{}\Big]_{ij} \nonumber\\
    &\phantom{\to\asix{\psi\phi}{i j a b c} +\frac{1}{2!3!}\sum_{\mathrm{perms}}\Bigg\{}- \frac{51}{4}\,\Big[Y_{c}  \rfive{\psi}{}^{\dagger}  \rfive{\psi}{}  Y^{\dagger}_a Y_{b}\Big]_{ij} - \frac{9}{2}\,\Big[Y_{c}  \rfive{\psi}{}^{\dagger}  Y_{a}  \rfive{\psi}{}^{\dagger}  Y_{b}\Big]_{ij} \nonumber\\
    &\phantom{\to\asix{\psi\phi}{i j a b c} +\frac{1}{2!3!}\sum_{\mathrm{perms}}\Bigg\{}- \rsix{\psi\phi D1}{i j d}\,\lambda_{a b c d} - \frac{1}{2}\rsix{\psi\phi D2}{i j d}\,\lambda_{a b c d} \nonumber\\
    &\phantom{\to\asix{\psi\phi}{i j a b c} +\frac{1}{2!3!}\sum_{\mathrm{perms}}\Bigg\{}+ \rsix{D\phi}{d e}\,Y_{i j}^{d}\,\lambda_{a b c e} + \rsix{\phi D}{d a b c}\,Y_{i j}^{d} \nonumber\\
    &\phantom{\to\asix{\psi\phi}{i j a b c} +\frac{1}{2!3!}\sum_{\mathrm{perms}}\Bigg\{}+ 6\,\Big[Y_{c}  \rsix{\phi\psi1}{a b}\Big]_{ij} - 6i\,\Big[Y_{c}  \rsix{\phi\psi2}{a b}\Big]_{ij} \nonumber\\
    &\phantom{\to\asix{\psi\phi}{i j a b c} +\frac{1}{2!3!}\sum_{\mathrm{perms}}\Bigg\{}- 3\,\Big[Y_{c}  \rsix{\psi\phi D2}{a}^\dagger  Y_{b}\Big]_{ij} + 3\,\Big[Y_{c}  Y^{\dagger}_b  \rsix{\psi\phi D2}{a}\Big]_{ij}\nonumber\\
    &\phantom{\to\asix{\psi\phi}{i j a b c} +\frac{1}{2!3!}\sum_{\mathrm{perms}}\Bigg\{}- 3\,\Big[Y_{c}  \rsix{\psi\phi D3}{a}^\dagger  Y_{b}\Big]_{ij} - 3\,\Big[Y_{c}  Y^{\dagger}_b  \rsix{\psi\phi D3}{a}\Big]_{ij}  \nonumber\\
    &\phantom{\to\asix{\psi\phi}{i j a b c} +\frac{1}{2!3!}\sum_{\mathrm{perms}}\Bigg\{}- 6\,\Big[Y_{c}  \rsix{\psi D}{}  Y^{\dagger}_a  Y_{b}\Big]_{ij}- 6\,\Big[Y_{c}  Y^{\dagger}_b  Y_{a}  \rsix{\psi D}{}\Big]_{ij}\Bigg\},   \\
    \asix{\psi\psi}{i j k l}&\to\asix{\psi\psi}{i j k l}+\frac{1}{4!}\sum_{\mathrm{perms}}\Bigg\{ - 4\,\rsix{\psi\psi}{j i k l}^\dagger - 8\,\rsix{\psi\psi}{l i k j}^\dagger \nonumber\\
    &\phantom{\to\asix{\psi\psi}{i j k l}+\frac{1}{4!}\sum_{\mathrm{perms}}\Bigg\{}- 2\rsix{\psi\phi D1}{i j a}\,Y_{k l a} - \rsix{\psi\phi D2}{i j a}\,Y_{k l a}\nonumber\\
    &\phantom{\to\asix{\psi\psi}{i j k l}+\frac{1}{4!}\sum_{\mathrm{perms}}\Bigg\{}+ 3\,\rsix{D\phi}{a b}\,Y_{i j a}\,Y_{k l b}\Bigg\}.
\end{align}
The coefficients not included in this list do not receive any contributions from redundant coefficients.

\section{Calculation of the beta functions \label{sec_beta_functions}}

\subsection{General formalism}

The procedure to compute the beta functions of an EFT is well known. Nevertheless, for the sake of clarity and for pedagogical reasons, we include here, and in Appendix~\ref{appendix_RGE}, a detailed derivation of this procedure.

Let us assume a UV effective Lagrangian of generic form
\begin{equation}
    \lag^{\mathrm{UV}}= \sum_{i} a_i \mathcal{O}_i, \label{ldiv}
\end{equation}
where $a_i$ denotes the Wilson coefficient of the operator $\mathcal{O}_i$ of arbitrary mass dimension and the sum runs over all the physical operators up to the desired dimension. This Lagrangian will induce UV divergences that, after subtraction of sub-divergences, can be matched to the full EFT (including non-trivial kinetic terms and, if the matching is performed off-shell, redundant operators). After canonical normalization and reduction to the physical basis, the divergent Lagrangian reads,
\begin{equation}
    \lag^{\mathrm{div}} = \sum_{j} a^\prime_j(a) ~\mathcal{O}_j,
\end{equation}
where we have denoted with a prime the WCs that parameterize the UV divergences (which are a function of the WCs in the UV Lagrangian, as explicitly shown). We use the background field method~\cite{Abbott:1980hw} and dimensional regularization to regularize the divergences, working in $d=4-2 \epsilon$ dimensions.
These divergences can be canceled by splitting the WCs into renormalized WCs and counterterms~\footnote{The fact that we have canonically normalized the divergent Lagrangian implies that our fields are already renormalized fields and we do not need to worry about wave-function renormalization factors.}
\begin{equation}
a_i^{\mathrm{bare}}= \mu^{(n_i-2)\epsilon} (a_i + \delta a_i),
\end{equation}
where the scale $\mu$ is introduced to ensure that the renormalized WCs have the same mass dimension in $d$ dimensions as the bare one in $d=4$, what fixes $n_i$ to the number of fields appearing in the operator $\mathcal{O}_i$.  
Since we have canonically normalized the divergent Lagrangian in Eq.\eqref{ldiv}, we then have,
\begin{equation}
\delta a_i = - a_i^\prime. \label{deltCieqCiprime}
\end{equation}

The corresponding counterterms fix the renormalization scale dependence of the renormalized WCs. At one loop they read (see Appendix~\ref{appendix_RGE} for a general derivation at all loop order)
\begin{equation}
\beta_i \equiv \mu \frac{\mathrm{d}a_i}{\mathrm{d}\mu} = -2 a_i^\prime. \label{beta_oneloop}
\end{equation}
We will report below the results in the form
\begin{equation}
    \dot{a}_i \equiv 16 \pi^2 \beta_i.
\end{equation}

\subsection{Matching of UV divergences}

In order to compute the beta functions of our general EFT we consider as UV boundary condition the physical Lagrangian
\begin{equation}
    \lag^{\mathrm{UV}}=\lag_{d\leq 4}+\lag_5^{\mathrm{phys}}+\lag_6^{\mathrm{phys}},
\end{equation}
with canonically normalized kinetic terms.

The one loop divergence to the kinetic terms read (recall that a dagger means transposition of fermionic indices besides complex conjugation),
\begin{align}
    \ol
    (a^\prime_{KF})_{AB} &=-\frac{11}{3} g^2 f^{ACD} f^{BCD}+\frac{1}{6}g^2 \tr{\theta^A \theta^B}
    +\frac{2}{3}g^2 \tr{t^A t^B}\nonumber \\
    &+2 g \Big[\tr{t^A m^\dagger_\psi \afive{\psi F}{B}}
    + \mathrm{h.c.} + (A\leftrightarrow B) \Big]
    \nonumber \\ &
    -4 \Big[\tr{m^\dagger_\psi \afive{\psi F}{A} 
    m^\dagger_\psi \afive{\psi F}{B}} + \mathrm{h.c.}  \Big]
    \nonumber \\ &
+ (m_\phi^2)_{ab} \Big[ 
2 \afive{\phi F}{ACa} \afive{\phi F}{BCb}
-\frac{5}{4} \afive{\phi \widetilde{F}}{ACa} \afive{\phi \widetilde{F}}{BCb}
+\asix{\phi F}{ABab}\Big]    
\label{aprimekf}, \\
    \ol
(a^\prime_{K\phi})_{ab} &= 
-2 g^2 \theta^A_{ac} \theta^A_{cb}
    +\frac{1}{2}\tr{Y_a Y_b^{\dagger}+Y_b Y_a^{\dagger} }
    -2 (m_\phi^2)_{cd} \asix{\phi D}{abcd} 
    , \\
    \ol
(a^\prime_{K\psi})_{ij} &= 
    g^2 \Big[t^A t^A\Big]_{ij}
    +\frac{1}{2} \Big[Y_a^\dagger Y_a\Big]_{ij}
-3 g \Big[t^A m^\dagger_\psi \afive{\psi F}{A}+ \mathrm{h.c.}\Big]_{ij}    
-12 \Big[\afive{\psi F}{A}^{\dagger} m_\psi m^\dagger_\psi \afive{\psi F}{A}\Big]_{ij}.
\end{align}

After canonical normalization, all one-loop divergences can be parametrized by our Green's effective Lagrangian. The corresponding divergent WCs are collected, in the Green's basis, in Appendix~\ref{matching:green}. These results were obtained using \texttt{MatchMakerEFT}~\cite{Carmona:2021xtq}, with a general model as well as with the use of specific toy models (whose WCs were computed with the help of \texttt{GroupMath}~\cite{Fonseca:2020vke}). For purely bosonic operators the results were also independently computed with functional methods (following closely \cite{Cohen:2020fcu,Fuentes-Martin:2023ljp} in particular) using special code developed by us, together with the \texttt{STrEAM} package \cite{Cohen:2020qvb}.

\subsection{Beta functions for the bosonic operators}

In the background field gauge we are using, the product
$g_A A_\mu^A$ does not renormalize and, therefore, the beta function of the gauge coupling is given directly by the renormalization of the gauge boson kinetic term,
\begin{equation}
\dot{g}_A = (a_{KF}^\prime)_{AB} g_B,    
\end{equation}
where $(a_{KF}^\prime)_{AB}$ is given in Eq.~\eqref{aprimekf} and, for clarity, we have reinstated the indices in the gauge couplings temporarily. As for the remaining WCs, 
reducing the divergent Lagrangian in the Green's basis to the physical one and using Eq. \eqref{beta_oneloop}, we obtain the following beta functions for the physical bosonic operators,~\footnote{In the following, we write the expressions in a way that makes their symmetries explicit. For instance $\alpha_{ab}=f_{ab}+(a \leftrightarrow b)$ means that the whole expression has to be added with $a$ and $b$ swapped, $\alpha_{ab}=f_{ab}+f_{ba}$ (even terms that are explicitly symmetric under this swapping). Similarly, when the corresponding symmetrization involves more than just transpositions, we will write a sum over permutations, possibly with a sign, that involves all the free indices unless otherwise explicitly indicated.}
\begin{align}
\dot{\eta}_a &=
 \kappa_{abc} (m_\phi^2)_{bc}
-2 \tr{Y_a m^\dagger_\psi m_\psi m^\dagger_\psi 
+ \mathrm{h.c.}}
\nonumber \\ 
&+ \eta_b \Big\{
-2 g^2 \theta^A_{ac} \theta^A_{cb} + \frac{1}{2} \tr{
Y_a Y^\dagger_b +Y_b Y^\dagger_a}
-2(m_\phi^2)_{cd} \asix{\phi D}{abcd}
\Big\}
\nonumber \\
&+2(m_\phi^2)_{ab}\afive{\phi F}{ABb} \afive{\phi F}{ABc} \eta_c
+2(m_\phi^2)_{ab}\afive{\phi \widetilde{F}}{ABb} \afive{\phi \widetilde{F}}{ABc}  \eta_c,\\
(\dot{m_\phi^2})_{ab}&=
\frac{1}{2}\kappa_{ac d}\kappa_{bcd} 
+\frac{1}{2} \lambda_{a b c d} (m_\phi^2)_{cd}
+2 g^2 (m_{\phi}^2)_{a c}\theta _ {b d}^{A}\theta _ {cd}^{A}
\nonumber\\&
+ \frac{1}{2} (m_{\phi}^2)_{ac}\, \tr{Y_{b} Y^\dagger_c+Y_{c} Y^\dagger_b} 
-\tr{m_\psi^\dagger Y_a m_\psi^\dagger Y_b
+m_\psi Y^{\dagger}_a m_\psi Y^{\dagger}_b}
-4~ \tr{ m_\psi^\dagger m_\psi Y^\dagger_a Y_b}
\nonumber \\  &
+6 g^2  \theta _{a c}^{A} \theta _{cd}^{B} \eta _{d} \afive{\phi F}{ABb}
+\bigg\{
\tr{m_\psi^\dagger m_\psi m_\psi^\dagger \afive{\psi\phi^2}{a b}} 
-\frac{1}{2} \tr{\afive{\psi\phi^2}{a b} Y^\dagger_c} \eta_c 
+ \mathrm{h.c.}
\bigg\}
\nonumber\\ &
+(m_{\phi }^2)_{ac} (m_{\phi }^2)_{bd} 
\afive{\phi F}{ABc} \afive{\phi F}{ABd}
+ (m_{\phi }^2)_{ac} (m_{\phi }^2)_{bd} 
\afive{\phi \widetilde{F}}{ABc} \afive{\phi \widetilde{F}}{ABd}
\nonumber \\ &
+ \kappa _{abc} \eta _{d} 
\afive{\phi F}{ABc} \afive{\phi F}{ABd}
+ \kappa _{abc} \afive{\phi \widetilde{F}}{ABc} 
\afive{\phi \widetilde{F}}{ABd} \eta _{d}
\nonumber\\&
-2 (m_{\phi }^2)_{bc} (m_{\phi }^2)_{de} 
\asix{\phi D}{acde}
-2 (m_{\phi }^2)_{cd} (m_{\phi }^2)_{de} \asix{\phi D}{a b c e}
\nonumber \\ &
-2\eta _{e} \kappa _{bcd} \asix{\phi D}{a cde}
+3 \eta _{c} \kappa _{cde}  \asix{\phi D}{a b de} 
+\bigg\{2\mathrm{i}\, \eta_c \tr{m_\psi^\dagger Y^a \asix{\phi\psi}{cb}} + \mathrm{h.c.}\bigg\}
+ (a\leftrightarrow b),
\\
\dot{\kappa}_{abc}=&\sum_{\mathrm{perm}}\Bigg\{ 
\frac{1}{2} \kappa _{c de} \lambda _{a bd e} 
+g^2  \theta _{a f}^{A} \theta _{d f}^{A} \kappa _{b cd}
+g^2  \theta _{a e}^{A} \theta _{b d}^{A} \kappa _{cde}
+\frac{1}{4} \kappa_{a b d} \tr{Y_c Y^{\dagger}_d+Y^{\dagger}_c Y_d}
\nonumber \\ &
\phantom{\sum_{\mathrm{perm}}\Bigg\{}
-2\,\tr{m_\psi Y^{\dagger}_a Y_{b}Y^{\dagger}_c+m_\psi^\dagger Y_a Y^\dagger_{b}Y_c} 
\nonumber \\ &
\phantom{\sum_{\mathrm{perm}}\Bigg\{}
+6 g^2 \theta _{a f}^A \theta _{d f}^B (m_{\phi }^2)_{b d} 
\afive{\phi F}{AB c} 
-\frac{1}{2} (m_{\phi}^2)_{c d} 
\tr{
\afive{\psi\phi^2}{a b} Y^\dagger_{d}
+\afive{\psi\phi^2}{a b}^\dagger Y_{d}
}
\nonumber \\ &
\phantom{\sum_{\mathrm{perm}}\Bigg\{}
+2\, 
\tr{
m_\psi^\dagger m_\psi Y^\dagger_a \afive{\psi\phi^2}{b c}
+m_\psi m_\psi^\dagger Y_a \afive{\psi\phi^2}{b c}^\dagger
} 
\nonumber\\&
\phantom{\sum_{\mathrm{perm}}\Bigg\{}
+ \tr{
m_\psi^\dagger Y_c m_\psi^\dagger \afive{\psi\phi^2}{a b}
+m_\psi Y_c^\dagger m_\psi \afive{\psi\phi^2}{a b}^\dagger
}
\nonumber\\&
\phantom{\sum_{\mathrm{perm}}\Bigg\{}
-\frac{1}{6} (m_{\phi }^2)_{d e} \afive{\phi }{abc de} 
\nonumber \\ &
\phantom{\sum_{\mathrm{perm}}\Bigg\{}
+ (m_{\phi }^2)_{a e} \kappa _{b cd} 
\afive{\phi F}{ABd}  \afive{\phi F}{ABe}
-\frac{1}{6}\eta _{g} \lambda _{a b cd} 
\afive{\phi F}{ABd} \afive{\phi F}{ABg}
\nonumber \\ &
\phantom{\sum_{\mathrm{perm}}\Bigg\{}
+10 g^2 \eta _{g} \afive{\phi F}{A B a} 
\afive{\phi F}{A C b} \theta _{ c e }^{C} \theta _{ g e }^{B}
+4 g^2 \eta _{d} 
\afive{\phi F}{A B a} \afive{\phi F}{A C d} 
\theta _{ b e }^{B} \theta _{ c e }^{C}
\nonumber \\ &
\phantom{\sum_{\mathrm{perm}}\Bigg\{}
-2g^2 \eta _{d} 
\afive{\phi \tilde{F}}{AC b} 
\afive{\phi \tilde{F}}{BC c} \theta _{a e}^{A} \theta _{d e}^{B}
+4 g^2 
\eta _{d} 
\afive{\phi \tilde{F}}{AC c} 
\afive{\phi \tilde{F}}{BC d} 
\theta _{a e}^{A} \theta _{b e}^{B}
\nonumber \\ &
\phantom{\sum_{\mathrm{perm}}\Bigg\{}
+\frac{1}{3}\eta _{e} \lambda _{a b c d} 
\afive{\phi \tilde{F}}{AB d} \afive{\phi \tilde{F}}{AB e}
+ (m_{\phi }^2)_{a e} \kappa _{b c d} 
\afive{\phi \tilde{F}}{AB d} 
\afive{\phi \tilde{F}}{AB e}
\nonumber \\ &
\phantom{\sum_{\mathrm{perm}}\Bigg\{}
+\frac{1}{6} 
\tr{
\afive{\psi\phi^2}{d a} \afive{\psi\phi^2}{b c}^\dagger
+\afive{\psi\phi^2}{d a}^\dagger \afive{\psi\phi^2}{b c}
} \eta_{d} 
\nonumber\\&
\phantom{\sum_{\mathrm{perm}}\Bigg\{}
+3g^2  \eta _{e} \theta _{a f}^{A} \theta _{e f}^{B} 
\asix{\phi F}{ABbc}
\nonumber \\ &
\phantom{\sum_{\mathrm{perm}}\Bigg\{}
-4 \kappa _{c d e} (m_{\phi }^2)_{d f} 
\asix{\phi D}{a b f e} 
+3 (m_{\phi }^2)_{b d} \kappa _{d e f} 
\asix{\phi D}{a c e f}
-\frac{2}{3} \kappa _{b c d} (m_{\phi }^2)_{e f} 
\asix{\phi D}{a d e f}
\nonumber\\ &
\phantom{\sum_{\mathrm{perm}}\Bigg\{}
+\frac{2}{3} \kappa _{b c d} (m_{\phi }^2)_{e f} 
\asix{\phi D}{a e d f}
-2 (m_{\phi }^2)_{b f} \kappa _{c d e} 
\asix{\phi D}{f a d e}
+ \eta _{d} 
\asix{\phi D}{e f a b} \lambda _{d c e f}
\nonumber \\ &
\phantom{\sum_{\mathrm{perm}}\Bigg\{}
+3g^2  \eta _{d} \theta _{a e}^{A} \theta _{b f}^{A} 
\asix{\phi D}{f e d c}
+3g^2  
\eta _{d} \theta _{a e}^{A} \theta _{f e}^{A} 
\asix{\phi D}{d f c b}
+3g^2  \eta _{f} \asix{\phi D}{e d c a} 
\theta _{b e}^{A} \theta _{f d}^{A}
\nonumber\\ &
\phantom{\sum_{\mathrm{perm}}\Bigg\{}
+
2\mathrm{i} (m_\phi^2)_{b d}
\tr{
\asix{\phi\psi}{d c} Y_a m_\psi^\dagger    
-m_\psi Y^\dagger_a    \asix{\phi\psi}{d c}
} 
\nonumber\\ &
\phantom{\sum_{\mathrm{perm}}\Bigg\{}
+\frac{1}{3} 
\tr{
m_\psi^\dagger m_\psi m_\psi^\dagger \asix{\psi\phi}{a b c}
+m_\psi m_\psi^\dagger m_\psi \asix{\psi\phi}{a b c}^\dagger
}
\nonumber\\ &
\phantom{\sum_{\mathrm{perm}}\Bigg\{ }
-\frac{1}{6} \,
\tr{
\asix{\psi\phi}{a b c} Y^\dagger_{d}
+\asix{\psi\phi}{a b c}^\dagger Y_{d}
} \eta_{d}
\Bigg\},
\\
\dot{\lambda}_{abcd}&=\sum_{\mathrm{perm}}\Bigg\{
\frac{3 g^4}{2} \theta_{af}^B \theta_{ce}^A \theta_{df}^A \theta_{be}^B
-\frac{3 g^2}{2} \lambda _{acfe} \theta_{be}^A \theta_{fd}^A
+\frac{1}{8} \lambda _{abef} \lambda _{cdef} 
\nonumber\\   &
\phantom{\sum_{\mathrm{perm}}\Bigg\{}
- \tr{Y^{\dagger }_a Y_c Y^{\dagger}_b Y_d} 
+\frac{1}{12} \tr{Y_d Y^{\dagger}_e+Y_d^\dagger Y_e}  \lambda_{abce}
\nonumber \\ &
\phantom{\sum_{\mathrm{perm}}\Bigg\{}
+ 6 g^2 \kappa_{efd} 
\afive{\phi F}{ABa} \theta_{fc}^A \theta_{eb}^B
-\frac{1}{6} \kappa _{aef}  \afive{\phi }{bcdef}
\nonumber\\ &
\phantom{\sum_{\mathrm{perm}}\Bigg\{}
-\frac{1}{4} \kappa _{bce} 
\tr{Y^{\dagger}_e  \afive{\psi \phi^2}{ad}
+Y_e  \afive{\psi \phi^2}{ad}^{\dagger}}
+2 \tr{
m_{\psi}^{\dagger } Y_{b} \,Y^{\dagger}_a  \afive{\psi \phi^2}{cd}
+m_{\psi} Y_{b}^{\dagger } \,Y_a  \afive{\psi \phi^2}{cd}^{\dagger}
}
\nonumber\\ &
\phantom{\sum_{\mathrm{perm}}\Bigg\{}
+ \tr{
Y_{b}^\dagger m_\psi Y_{a}^{\dagger } \afive{\psi \phi^2}{cd} 
+Y_{b} m_\psi^{\dagger } Y_{a} \afive{\psi \phi^2}{cd}^\dagger 
}
\nonumber \\ &
\phantom{\sum_{\mathrm{perm}}\Bigg\{}
+ 4 g^2 (m_{\phi }^2)_{ae} 
\afive{\phi F}{ABd} \afive{\phi F}{ACe} 
\theta_{bf}^B \theta_{cf}^C
+10 g^2 (m_{\phi }^2)_{af} \afive{\phi F}{ABD} 
\afive{\phi F}{ACc} \theta_{be}^B \theta_{fe}^C
\nonumber\\ &
\phantom{\sum_{\mathrm{perm}}\Bigg\{}
+ 4 g^2 (m_{\phi }^2)_{ae} \afive{\phi \tilde{F}}{ABe} 
\afive{\phi \tilde{F}}{ACd} \theta_{bf}^B \theta_{cf}^C
- 2 g^2 (m_{\phi }^2)_{af} 
\afive{\phi \tilde{F}}{ABc} \afive{\phi \tilde{F}}{ACd} 
\theta_{be}^B \theta_{fe}^C
\nonumber\\&
\phantom{\sum_{\mathrm{perm}}\Bigg\{}
+\frac{1}{4} \kappa _{abe} \kappa _{cdf} 
\afive{\phi F}{ABe} \afive{\phi F}{ABf}
+\frac{1}{4} \kappa _{abe} \kappa _{cdf} 
\afive{\phi \tilde{F}}{ABe} 
\afive{\phi \tilde{F}}{ABf}
\nonumber\\ &
\phantom{\sum_{\mathrm{perm}}\Bigg\{}
+\frac{1}{3} (m_{\phi }^2)_{af} 
\afive{\phi F}{ABe} 
\afive{\phi F}{ABf} \lambda _{bcde}
+\frac{1}{3} (m_{\phi }^2)_{af} 
\afive{\phi \tilde{F}}{ABe} \afive{\phi \tilde{F}}{ABf} 
\lambda _{bcde}
\nonumber\\ &
\phantom{\sum_{\mathrm{perm}}\Bigg\{}
+\frac{3}{2} \kappa _{bce} \kappa _{efg} 
\asix{\phi D}{adfg}
- 2 \kappa _{aef} \kappa _{bgf} 
\asix{\phi D}{cdeg}
- \kappa _{aef} \kappa _{bcg} 
\asix{\phi D}{gdef}
\nonumber\\&
\phantom{\sum_{\mathrm{perm}}\Bigg\{}
+\frac{1}{6} (m_{\phi }^2)_{ae} \tr{
\afive{\psi \phi^2}{ed} \afive{\psi \phi^2}{bc}^\dagger}
+\frac{1}{6} (m_{\phi }^2)_{ae} \tr{\afive{\psi \phi^2}{cd} 
\afive{\psi \phi^2}{eb}^\dagger}
\nonumber\\ &
\phantom{\sum_{\mathrm{perm}}\Bigg\{}
- \tr{
m_{\psi }^{\dagger } m_{\psi }  
\afive{\psi \phi^2}{ab}^\dagger
\afive{\psi \phi^2}{cd} 
}
\nonumber\\ &
\phantom{\sum_{\mathrm{perm}}\Bigg\{}
-\frac{1}{4} \tr{
m_{\psi } \afive{\psi \phi^2}{ab}^\dagger 
m_{\psi } \afive{\psi \phi^2}{cd}^\dagger 
+m_{\psi }^\dagger \afive{\psi \phi^2}{ab} 
m_{\psi }^\dagger \afive{\psi \phi^2}{cd}
}
\nonumber \\ &
\phantom{\sum_{\mathrm{perm}}\Bigg\{}
-3 g^2 (m_{\phi }^2)_{ef} \theta_{ea}^A \theta_{bf }^B 
\asix{\phi F}{ABcd}
- 12 g^2 (m_{\phi }^2)_{ae} \theta_{cf}^A \theta_{eg}^A 
\asix{\phi D}{bfgd}
\nonumber\\ &
\phantom{\sum_{\mathrm{perm}}\Bigg\{}
-2 (m_{\phi }^2)_{ef} \lambda _{bcge} 
\asix{\phi D}{adgf}
+\frac{2}{3} (m_{\phi }^2)_{ef} \lambda _{abcg} 
\asix{\phi D}{degf}
\nonumber\\ &
\phantom{\sum_{\mathrm{perm}}\Bigg\{}
+ (m_{\phi }^2)_{ae} \lambda _{ebfg} 
\asix{\phi D}{fgcd}
-\frac{1}{24} (m_{\phi }^2)_{ef} 
\,\asix{\phi }{abcdef}
\nonumber\\ &
\phantom{\sum_{\mathrm{perm}}\Bigg\{}
-\mathrm{i} \kappa _{bce}\, 
\tr{
Y^{\dagger}_a m_{\psi } \asix{\phi \psi }{ed}
- \asix{\phi \psi }{ed} m_\psi^\dagger Y_a
} 
\nonumber\\ &
\phantom{\sum_{\mathrm{perm}}\Bigg\{}
+\frac{1}{3} 
\tr{
m_{\psi }^{\dagger } Y_{a} m_{\psi }^{\dagger } \asix{\psi \phi }{bcd} 
+m_{\psi } Y_{a}^{\dagger } m_{\psi } \asix{\psi \phi }{bcd}^{\dagger } 
}
\nonumber\\ &
\phantom{\sum_{\mathrm{perm}}\Bigg\{}
-\frac{1}{6}  (m_{\phi }^2)_{ae} 
\tr{
Y^{\dagger}_e\asix{\psi \phi }{bcd}
+Y_e\asix{\psi \phi }{bcd}^{\dagger}
}
\nonumber\\   &
\phantom{\sum_{\mathrm{perm}}\Bigg\{}
+\frac{2}{3} 
\tr{
Y^{\dagger}_a m_\psi m_\psi^\dagger \asix{\psi \phi }{bcd} 
+Y_a m_\psi^\dagger m_\psi \asix{\psi \phi }{bcd}^\dagger 
}
\Bigg\},
\\
\afived{\phi}{abcde}&=
\sum_{\mathrm{perm}}\Bigg\{
- 6 g^4 \afive{\phi F}{ABe} \theta_{af}^A \theta_{bf}^C 
\theta_{cg}^B \theta_{dg}^C
+ 3g^2 \afive{\phi F}{ABe} \lambda _{bcgf} 
\theta_{af}^A \theta_{gd}^B
\nonumber \\ &
\phantom{\sum_{\mathrm{perm}}\Bigg\{}
+\frac{g^2}{2} \theta_{af}^A \theta_{bg}^A 
\afive{\phi}{cdefg}
+\frac{1}{12}\lambda _{abfg} 
\afive{\phi}{cdefg}
\nonumber \\ &
\phantom{\sum_{\mathrm{perm}}\Bigg\{}
-  \tr{
Y^{\dagger}_b Y_a  Y^{\dagger}_c \afive{\psi \phi^2}{de}
+Y_b Y^\dagger_a  Y_c \afive{\psi \phi^2}{de}^\dagger
}
\nonumber \\ &
\phantom{\sum_{\mathrm{perm}}\Bigg\{}
+\frac{1}{12}\lambda _{bcdf} 
\tr{
Y^\dagger_f \afive{\psi \phi^2}{ae} 
+Y_f \afive{\psi \phi^2}{ae}^\dagger 
}
\nonumber \\ &
\phantom{\sum_{\mathrm{perm}}\Bigg\{}
+\frac{1}{48} 
\tr{
Y_f Y^\dagger_a 
+Y_f^\dagger Y_a 
}
\afive{\phi}{bcdef} 
\nonumber \\ &
\phantom{\sum_{\mathrm{perm}}\Bigg\{}
- 2 g^2 \kappa _{aef} 
\afive{\phi F}{ABd} 
\afive{\phi F}{ACf} \theta_{bg}^B \theta_{cg}^C
-5 g^2 \kappa _{aeg} 
\afive{\phi F}{ABd} \afive{\phi F}{ACc} 
\theta_{bf}^B \theta_{gf}^C
\nonumber \\ &
\phantom{\sum_{\mathrm{perm}}\Bigg\{}
-2 g^2 \kappa _{aef} \afive{\phi\tilde{F}}{ABf} 
\afive{\phi\tilde{F}}{ACd} \theta_{bg}^B \theta_{cg}^C
+g^2 \kappa _{aeg} 
\afive{\phi\tilde{F}}{ABc} \afive{\phi\tilde{F}}{ACd} 
\theta_{bf}^B \theta_{gf}^C
\nonumber \\ &
\phantom{\sum_{\mathrm{perm}}\Bigg\{}
-\frac{1}{6}\kappa _{abf} 
\afive{\phi F}{ABf} \afive{\phi F}{ABg} \lambda _{cdeg}
-\frac{1}{6}\kappa _{abf} 
\afive{\phi\tilde{F}}{ABf} \afive{\phi\tilde{F}}{ABg} 
\lambda _{cdeg}
\nonumber \\ &
\phantom{\sum_{\mathrm{perm}}\Bigg\{}
+\frac{1}{2} 
\tr{
m_\psi^\dagger \afive{\psi \phi^2}{bc} Y^\dagger_a \afive{\psi \phi^2}{de} 
+m_\psi \afive{\psi \phi^2}{bc}^\dagger Y_a \afive{\psi \phi^2}{de}^\dagger 
}
\nonumber\\ &
\phantom{\sum_{\mathrm{perm}}\Bigg\{}
+ \tr{
m_\psi  Y_a^\dagger \afive{\psi \phi^2}{bc} \afive{\psi \phi^2}{de}^\dagger
+m_\psi^\dagger  Y_a \afive{\psi \phi^2}{bc}^\dagger \afive{\psi \phi^2}{de}
}
\nonumber\\ &
\phantom{\sum_{\mathrm{perm}}\Bigg\{}
-\frac{1}{6}\kappa _{aef} \tr{
\afive{\psi \phi^2}{fd} \afive{\psi \phi^2}{bd}^\dagger 
}
\nonumber \\ &
\phantom{\sum_{\mathrm{perm}}\Bigg\{}
-\frac{3 g^2}{2} \kappa _{aeh} \theta_{hg}^A \theta_{bg}^B 
\asix{\phi F}{ABcd}
+6 g^2 \kappa _{aef} \theta_{cg}^A \theta_{fh}^A 
\asix{\phi D}{bghd}
\nonumber \\ &
\phantom{\sum_{\mathrm{perm}}\Bigg\{}
-\frac{1}{2} \kappa _{fgh} \lambda _{bcdf} 
\asix{\phi D}{aegh}
+\frac{1}{3} \kappa _{afg} \lambda _{bcdh} 
\asix{\phi D}{hefg}
\nonumber \\ &
\phantom{\sum_{\mathrm{perm}}\Bigg\{}
+2 \kappa _{afg} \lambda _{bchg} 
\asix{\phi D}{dehf}
-\frac{1}{2} \kappa _{aef} \lambda _{fbgh} 
\asix{\phi D}{ghcd}
\nonumber \\ &
\phantom{\sum_{\mathrm{perm}}\Bigg\{}
+\frac{1}{24}\kappa_{afg} 
\asix{\phi}{bcdefg}
\nonumber\\ &
\phantom{\sum_{\mathrm{perm}}\Bigg\{}
-\frac{2}{3} 
\tr{
m_\psi^{\dagger } Y_{a} Y^{\dagger}_b \asix{\psi \phi }{cde} 
+m_\psi Y_{a}^{\dagger } Y_b \asix{\psi \phi }{cde}^{\dagger} 
}
\nonumber \\ &
\phantom{\sum_{\mathrm{perm}}\Bigg\{}
-\frac{1}{3} 
\tr{ 
Y^\dagger_a m_\psi Y^\dagger_b \asix{\psi \phi }{cde} 
+Y_a m_\psi^\dagger Y_b \asix{\psi \phi }{cde}^\dagger 
}
\nonumber \\ &
\phantom{\sum_{\mathrm{perm}}\Bigg\{}
+\frac{1}{12}\kappa _{aef} 
\tr{
Y^\dagger_f \asix{\psi \phi }{bcd} 
+Y_f \asix{\psi \phi }{bcd}^\dagger 
}
\nonumber \\ &
\phantom{\sum_{\mathrm{perm}}\Bigg\{}
+\frac{\mathrm{i}}{3}  \lambda _{bcdf} 
\tr{ 
Y^\dagger_a m_\psi \asix{\phi \psi }{fe} 
+ \asix{\phi \psi}{fe} m_\psi^\dagger Y_a
}
\Bigg\},
    \\
\afived{\phi F}{ABa} &=
-4 g^2 f^{ACD} f^{BED} \afive{\phi F}{ECa}
+\frac{g^2}{3} f^{CDE} f^{ADE} \afive{\phi F}{ACa} 
\nonumber \\ &
+2g^2 \theta^{C}_{ac}\theta^B_{cb}\afive{\phi F}{ACb}
-g^2 \theta^{C}_{ac}\theta^C_{cb}\afive{\phi F}{ABb}
+\frac{g^2}{6} \theta^{B}_{bc}\theta^C_{cb}\afive{\phi F}{ACa}
+\frac{2g^2}{3} \tr{t^B t^C}
\afive{\phi F}{ACa}
\nonumber \\ &
+2g \tr{ t^B \afive{\psi F}{A}^\dagger Y_a
+Y_a^\dagger \afive{\psi F}{A}t^B}
+\frac{1}{4}
\tr{Y_a Y_b^\dagger + Y_b Y_a^\dagger}
\afive{\phi F}{ABb}
\nonumber \\ &
+\kappa_{abc} \afive{\phi F}{ACb} \afive{\phi F}{BCc}
-\kappa_{abc} \afive{\phi \widetilde{F}}{ACb} \afive{\phi \widetilde{F}}{BCc}
\nonumber \\ &
+\Big\{2g \afive{\phi F}{ACa} 
\tr{\afive{\psi F}{B} t^C m_\psi^\dagger
+\afive{\psi F}{C} t^B m_\psi^\dagger
}
+4 \tr{
\afive{\psi F}{A}Y_a^\dagger \afive{\psi F}{B} m_\psi^\dagger}
+\mathrm{h.c.}\Big\}
\nonumber \\ &
+\frac{1}{2} \kappa_{abc} \asix{\phi F}{ABbc}
+2 g \tr{\asix{\psi F}{Aa} t^B m_\psi^\dagger
+m_\psi t^B \asix{\psi F}{Aa}^\dagger }
+(A\leftrightarrow B),
\\
\afived{\phi \widetilde{F}}{ABa} &=
-4g^2 f^{ACD}f^{BED} \afive{\phi \widetilde{F}}{ECa}
+\frac{g^2}{3} f^{CDE} f^{ADE} \afive{\phi \widetilde{F}}{ACa} 
\nonumber \\ &
+2g^2 \theta^{C}_{ac}\theta^B_{cb}\afive{\phi \widetilde{F}}{ACb}
-g^2 \theta^{C}_{ac}\theta^C_{cb}\afive{\phi \widetilde{F}}{ABb}
+\frac{g^2}{6} \theta^{B}_{bc}\theta^C_{cb}\afive{\phi \widetilde{F}}{ACa}
+\frac{2g^2}{3} \tr{t^B t^C}
\afive{\phi \widetilde{F}}{ACa}
\nonumber \\ &
-2\mathrm{i} g \tr{\afive{\psi F}{A} t^B Y_a^\dagger
-Y_a t^B\afive{\psi F}{A}^\dagger }
+\frac{1}{4} \afive{\phi \widetilde{F}}{ABb} 
\tr{
Y_a Y_b^\dagger+Y_b Y_a^\dagger
}
\nonumber \\ &
+2\kappa_{abc} \afive{\phi F}{ACb} \afive{\phi \widetilde{F}}{BCc}
\nonumber \\ &
+\Big\{
2g \afive{\phi \widetilde{F}}{ACa} 
\tr{\afive{\psi F}{B} t^C m_\psi^\dagger
+\afive{\psi F}{C} t^B m_\psi^\dagger
} 
-4\mathrm{i} \tr{
\afive{\psi F}{A}Y_a^\dagger \afive{\psi F}{B} m_\psi^\dagger }
+ \mathrm{h.c.} \Big\}
\nonumber \\ &
+\frac{1}{2} \kappa_{abc} \asix{\phi \widetilde{F}}{ABbc}
-2\mathrm{i} g \tr{\asix{\psi F}{Aa} t^B m_\psi^\dagger
-m_\psi t^B \asix{\psi F}{Aa}^\dagger }
+(A\leftrightarrow B),
\\
\asixd{3F}{ABC} &=
\sum_{\pi(ABC)} (-1)^{P_\pi} \Big\{
2 g f^{DEA} \afive{\phi F}{DCa} \afive{\phi F}{EBa}
-2 g f^{DEA} \afive{\phi \widetilde{F}}{DCa} \afive{\phi \widetilde{F}}{EBa}
\nonumber \\
&
\phantom{\sum_{\pi(ABC)} (-1)^{P_\pi} \Big\{}
- \frac{16}{9} \mathrm{i} \tr{\afive{\psi F}{B} t^C \afive{\psi F}{A}^\dagger}
+\frac{g^2}{6} \asix{3F}{ABD} f^{CEF} f^{DEF}
\nonumber \\
&
\phantom{\sum_{\pi(ABC)} (-1)^{P_\pi} \Big\{}
+\frac{1}{12} g^2 \asix{3F}{BCD}\tr{\theta^A \theta^D}
+\frac{1}{3} g^2 \asix{3F}{BCD}\tr{t^A t^D}
\Big\},
\\
\asixd{3\widetilde{F}}{ABC} &=
\sum_{\pi(ABC)} (-1)^{P_\pi} \Big\{
-4 g f^{ADE} \afive{\phi F}{BDa}\afive{\phi \widetilde{F}}{CEa}
+\frac{g^2}{6} \asix{3\widetilde{F}}{ABD} f^{CEF} f^{DEF}
\nonumber \\
&
\phantom{\sum_{\pi(ABC)} (-1)^{P_\pi} \Big\{}
+\frac{1}{12} g^2 \asix{3\widetilde{F}}{BCD}\tr{\theta^A \theta^D}
+\frac{1}{3} g^2 \asix{3\widetilde{F}}{BCD}\tr{t^A t^D}
\Big\},
\\
\asixd{\phi D}{abcd}=&\Bigg\{ 
    -\frac{1}{9}g^2 \Big(\theta^A_{db}\theta^B_{ca} + \theta^A_{da}\theta^B_{cb}\Big)\afive{\phi F}{CAe}\afive{\phi F}{CBe}
    \nonumber\\
    &-\frac{1}{3}g^2 \Big(
    \theta^A_{be}\theta^B_{ae} + \theta^A_{ae}\theta^B_{be}
    \Big)
    \afive{\phi F}{CAd}\afive{\phi F}{CBc}
    \nonumber\\
    &+\frac{1}{6}g^2 \Big(
    \theta^A_{ce}\theta^B_{ae} + \theta^A_{ae}\theta^B_{ce}
    \Big)
    \afive{\phi F}{CAd}\afive{\phi F}{CBb}
    \nonumber\\
    &+\frac{1}{6}g^2 \Big(
    \theta^A_{de}\theta^B_{ae} + \theta^A_{ae}\theta^B_{de}
    \Big)
    \afive{\phi F}{CAb}\afive{\phi F}{CBc}
    \nonumber\\
    &+\frac{1}{6}g^2 \Big(
    \theta^A_{ce}\theta^B_{be} + \theta^A_{be}\theta^B_{ce}
    \Big)
    \afive{\phi F}{CAd}\afive{\phi F}{CBa}
    \nonumber\\
    &+\frac{1}{6}g^2 \Big(
    \theta^A_{de}\theta^B_{be} + \theta^A_{be}\theta^B_{de}
    \Big)
    \afive{\phi F}{CAa}\afive{\phi F}{CBc}
    \nonumber\\
    &-\frac{1}{3}g^2 \Big(
    \theta^A_{de}\theta^B_{ce} + \theta^A_{ce}\theta^B_{de}
    \Big)
    \afive{\phi F}{CAa}\afive{\phi F}{CBb}
    \nonumber\\
    &
    -\frac{17}{18}\mathrm{i}g^2 \theta^A_{ca}f^{BCA} \afive{\phi F}{DBd}\afive{\phi F}{DCb}
    -\frac{17}{18}\mathrm{i}g^2 \theta^A_{cb}f^{BCA} \afive{\phi F}{DBd}\afive{\phi F}{DCa}
    \nonumber\\
    &-\frac{17}{18}\mathrm{i}g^2 \theta^A_{da}f^{BCA} \afive{\phi F}{DBc}\afive{\phi F}{DCb}
    -\frac{17}{18}\mathrm{i}g^2 \theta^A_{db}f^{BCA} \afive{\phi F}{DBc}\afive{\phi F}{DCa}
    \nonumber\\
    &-\frac{1}{9}g^2 \Big(\theta^A_{db}\theta^B_{ca} + \theta^A_{da}\theta^B_{cb}\Big)\afive{\phi \widetilde{F}}{CAe}\afive{\phi \widetilde{F}}{CBe}\nonumber\\
    &-\frac{1}{3}g^2 \Big(
    \theta^A_{be}\theta^B_{ae} + \theta^A_{ae}\theta^B_{be}
    \Big)
    \afive{\phi \widetilde{F}}{CAd}\afive{\phi \widetilde{F}}{CBc}
    \nonumber\\
    &+\frac{1}{6}g^2 \Big(
    \theta^A_{ce}\theta^B_{ae} + \theta^A_{ae}\theta^B_{ce}
    \Big)
    \afive{\phi \widetilde{F}}{CAd}\afive{\phi \widetilde{F}}{CBb}
    \nonumber\\
    &+\frac{1}{6}g^2 \Big(
    \theta^A_{de}\theta^B_{ae} + \theta^A_{ae}\theta^B_{de}
    \Big)
    \afive{\phi \widetilde{F}}{CAb}\afive{\phi \widetilde{F}}{CBc}
    \nonumber\\
    &+\frac{1}{6}g^2 \Big(
    \theta^A_{ce}\theta^B_{be} + \theta^A_{be}\theta^B_{ce}
    \Big)
    \afive{\phi \widetilde{F}}{CAd}\afive{\phi \widetilde{F}}{CBa}
    \nonumber\\
    &+\frac{1}{6}g^2 \Big(
    \theta^A_{de}\theta^B_{be} + \theta^A_{be}\theta^B_{de}
    \Big)
    \afive{\phi \widetilde{F}}{CAa}\afive{\phi \widetilde{F}}{CBc}
    \nonumber\\
    &-\frac{1}{3}g^2 \Big(
    \theta^A_{de}\theta^B_{ce} + \theta^A_{ce}\theta^B_{de}
    \Big)
    \afive{\phi \widetilde{F}}{CAa}\afive{\phi \widetilde{F}}{CBb}
    \nonumber\\
    &
    -\frac{17}{18}\mathrm{i}g^2 \theta^A_{ca}f^{BCA} \afive{\phi \widetilde{F}}{DBd}\afive{\phi \widetilde{F}}{DCb}
    -\frac{17}{18}\mathrm{i}g^2 \theta^A_{cb}f^{BCA} \afive{\phi \widetilde{F}}{DBd}\afive{\phi \widetilde{F}}{DCa}
    \nonumber\\
    &-\frac{17}{18}\mathrm{i}g^2 \theta^A_{da}f^{BCA} \afive{\phi \widetilde{F}}{DBc}\afive{\phi \widetilde{F}}{DCb}
    -\frac{17}{18}\mathrm{i}g^2 \theta^A_{db}f^{BCA} \afive{\phi \widetilde{F}}{DBc}\afive{\phi \widetilde{F}}{DCa}
    \nonumber\\
    &
    -\frac{1}{9}g^2\Big(
    \theta^A_{ca}\theta^B_{db}
    +\theta^A_{cb}\theta^B_{da}
    +\theta^A_{da}\theta^B_{cb}
    +\theta^A_{ca}\theta^B_{db}
    \Big)\tr{\afive{\psi F}{A}\afive{\psi F}{B}^\dagger}
    \nonumber\\
    &+\frac{1}{18} \tr{\afive{\psi\phi 2}{c  d} \afive{\psi\phi 2}{a  b}^\dagger} - \frac{1}{36} \tr{\afive{\psi\phi 2}{b  d} \afive{\psi\phi 2}{ a  c}^\dagger} - \frac{1}{36} \tr{\afive{\psi\phi 2}{c  b} \afive{\psi\phi 2}{a  d}^\dagger}\nonumber\\
    &-\frac{1}{36} \tr{\afive{\psi\phi 2}{ a  d} \afive{\psi\phi 2}{c  b}^\dagger} + \frac{1}{18} \tr{\afive{\psi\phi 2}{ a  b} \afive{\psi\phi 2}{c  d}^\dagger} - \frac{1}{36}\tr{\afive{\psi\phi 2}{ c  a} \afive{\psi\phi 2}{ d  b}^\dagger}\nonumber\\
    &-\frac{5}{3}g^2\theta^A_{ae}\theta^A_{bf}\Big( \asix{\phi D}{cedf}+\asix{\phi D}{cfed}\Big)
    -\frac{5}{3}g^2\theta^A_{ce}\theta^A_{df}\Big( \asix{\phi D}{aebf}+\asix{\phi D}{afeb}\Big)
    \nonumber\\
    &
    +\frac{2}{3}g^2\theta^A_{ae}\theta^A_{cf}\asix{\phi D}{bedf}
    +\frac{2}{3}g^2\theta^A_{ae}\theta^A_{df}\asix{\phi D}{becf}
    +\frac{2}{3}g^2\theta^A_{be}\theta^A_{cf}\asix{\phi D}{aedf}
    +\frac{2}{3}g^2\theta^A_{be}\theta^A_{df}\asix{\phi D}{aecf}
    \nonumber\\
    &
    +g^2\theta^A_{ae}\theta^A_{cf}\asix{\phi D}{bfed}
    +g^2\theta^A_{be}\theta^A_{cf}\asix{\phi D}{afed}
    +g^2\theta^A_{ae}\theta^A_{df}\asix{\phi D}{bfec}
    +g^2\theta^A_{be}\theta^A_{df}\asix{\phi D}{afec}
    \nonumber\\
    &+\frac{1}{6}g^2\theta^A_{ae}\theta^A_{ef}\Big(\asix{\phi D}{bcfd}+\asix{\phi D}{bdfc}\Big)
    +\frac{1}{6}g^2\theta^A_{be}\theta^A_{ef}\Big(\asix{\phi D}{acfd}+\asix{\phi D}{adfc}\Big)
    \nonumber\\
    &-\frac{1}{6}g^2 \theta^A_{ce}\theta^A_{ef} \asix{\phi D}{abfd}
    -\frac{1}{6}g^2 \theta^A_{de}\theta^A_{ef} \asix{\phi D}{abfc}\nonumber\\
    &+\frac{1}{9}g^2\theta^A_{da}\theta^A_{ef}\asix{\phi D}{becf}
    +\frac{1}{9}g^2\theta^A_{db}\theta^A_{ef}\asix{\phi D}{aecf}
    +\frac{1}{9}g^2\theta^A_{ca}\theta^A_{ef}\asix{\phi D}{bedf}
    +\frac{1}{9}g^2\theta^A_{cb}\theta^A_{ef}\asix{\phi D}{aedf}\nonumber\\
    &+\frac{1}{3} \lambda_{abfe} \asix{\phi D}{fecd} -\frac{1}{6} \lambda_{acfe}\asix{\phi D}{febd}
    -\frac{1}{6} \lambda_{adfe}\asix{\phi D}{febc}\nonumber\\
    &-\frac{1}{6} \lambda_{bcfe}\asix{\phi D}{fead}
    -\frac{1}{6} \lambda_{bdfe}\asix{\phi D}{feac}
    +\frac{1}{3} \lambda_{cdfe}\asix{\phi D}{feab}\nonumber\\
    &-\frac{\mathrm{i}}{9}\theta^A_{db} \tr{t^A\asix{\phi\psi}{ac}}
    -\frac{\mathrm{i}}{9}\theta^A_{cb} \tr{t^A\asix{\phi\psi}{ad}}
    -\frac{\mathrm{i}}{9}\theta^A_{da} \tr{t^A\asix{\phi\psi}{bc}}
    -\frac{\mathrm{i}}{9}\theta^A_{ca} \tr{t^A\asix{\phi\psi}{bd}}
    \nonumber\\
    &+\frac{\mathrm{i}}{6}\tr{\asix{\phi\psi}{bd} Y_a^\dagger Y_c}
    +\frac{\mathrm{i}}{6}\tr{\asix{\phi\psi}{bc} Y_a^\dagger Y_d}
    +\frac{\mathrm{i}}{6}\tr{\asix{\phi\psi}{ad} Y_b^\dagger Y_c}\nonumber\\
    &+\frac{\mathrm{i}}{6}\tr{\asix{\phi\psi}{ac} Y_b^\dagger Y_d}
    -\frac{\mathrm{i}}{6}\tr{\asix{\phi\psi}{bd} Y_c^\dagger Y_a}
    -\frac{\mathrm{i}}{6}\tr{\asix{\phi\psi}{ad} Y_c^\dagger Y_b}\nonumber\\
    &-\frac{\mathrm{i}}{6}\tr{\asix{\phi\psi}{bc} Y_d^\dagger Y_a}
    -\frac{\mathrm{i}}{6}\tr{\asix{\phi\psi}{ac} Y_d^\dagger Y_b}\nonumber\\
    &+\Big\{-\frac{1}{9}\asix{\phi D}{bced} \tr{Y_e Y^\dagger_a}
    -\frac{1}{9}\asix{\phi D}{bdec} \tr{Y_e Y^\dagger_a}
    -\frac{1}{9}\asix{\phi D}{aced} \tr{Y_e Y^\dagger_b}\nonumber\\
    &-\frac{1}{9}\asix{\phi D}{adec} \tr{Y_e Y^\dagger_b}
    +\frac{1}{9}\asix{\phi D}{abed} \tr{Y_e Y^\dagger_c}
    +\frac{1}{9}\asix{\phi D}{abec} \tr{Y_e Y^\dagger_d}+\mathrm{h.c.}\Big\}\nonumber\\
    &+\left(ab\leftrightarrow cd\right)-\frac{1}{2}\left(a\leftrightarrow c\right)-\frac{1}{2}\left(a\leftrightarrow d\right)-\frac{1}{2}\left(b\leftrightarrow c\right)-\frac{1}{2}\left(b\leftrightarrow d\right)\Bigg\},\\
    \asixd{\phi F}{ABab}&=\sum_{\mathrm{perm}}\Bigg\{ 
    g^2\theta^C_{bd}\theta^D_{ad}\afive{\phi\tilde{F}}{ACc}\afive{\phi\tilde{F}}{BDc}
    +5g^2\theta^D_{cd}\theta^A_{ad}\afive{\phi\tilde{F}}{BCc}\afive{\phi\tilde{F}}{CDb}\nonumber\\
    &\phantom{=\sum_{\mathrm{perm}}\Bigg\{}
    -g^2\theta^C_{bd}\theta^D_{ad}\afive{\phi F}{ACc}\afive{\phi F}{BDc}
    -4g^2\theta^D_{cd}\theta^A_{ad}\afive{\phi F}{BCc}\afive{\phi F}{CDb}\nonumber\\
    &\phantom{=\sum_{\mathrm{perm}}\Bigg\{}
    +g^2\theta^D_{bc}\theta^D_{ad}\afive{\phi F}{ACc}\afive{\phi F}{BCd}\nonumber\\
    &\phantom{=\sum_{\mathrm{perm}}\Bigg\{}
    +\mathrm{i}g^2 \theta^D_{cb}f^{ECB}\afive{\phi\tilde{F}}{AEc}\afive{\phi\tilde{F}}{CDa}
    +\mathrm{i}g^2 \theta^E_{ca}f^{DCE}\afive{\phi\tilde{F}}{ACb}\afive{\phi\tilde{F}}{BDc}
    \nonumber\\
    &\phantom{=\sum_{\mathrm{perm}}\Bigg\{}+\frac{1}{2} \lambda_ {abdc} \afive{\phi F}{ACc} \afive{\phi F}{BCd}-\frac{1}{2} \lambda_{abdc} \afive{\phi\tilde{F}}{ACc} \afive{\phi\tilde{F}}{BCd}\nonumber\\
    &\phantom{=\sum_{\mathrm{perm}}\Bigg\{}+\Big\{-\frac{1}{4}\afive{\phi F}{ABc}\tr{\afive{\psi\phi 2}{ab}^\dagger Y_c}+ i \afive{\phi F}{ABc}\tr{m_\psi^\dagger Y_a \asix{\phi\psi}{cb}} + \mathrm{h.c.}\Big\}\nonumber\\
    &\phantom{=\sum_{\mathrm{perm}}\Bigg\{}-\frac{3}{2} g^2 \theta^{C}_{ad} \theta^{C}_{dc} \asix{\phi F}{ABcb}+\frac{1}{12} g^2 \theta^{A}_{cd} \theta^{C}_{dc} \asix{\phi F}{CBab}
    -2 g^2 \theta^{A}_{cd} \theta^{C}_{ad} \asix{\phi F}{BCbc}\nonumber\\
    &\phantom{=\sum_{\mathrm{perm}}\Bigg\{}-\frac{11}{6} g^2 f^{DCB} f^{DEA} \asix{\phi F}{CEab}
    +\frac{4}{3} \mathrm{i} g^2 \theta^{D}_{ca} f^{DCB} \asix{\phi F}{CAcb}\nonumber\\
    &\phantom{=\sum_{\mathrm{perm}}\Bigg\{}
    - \mathrm{i}  g^3 \theta^{C}_{ad} \theta^{A}_{cb} \theta^{D}_{dc}\asix{3F}{CBD} - 3 g^3 \theta^{C}_{ad}  \theta^{E}_{db} f^{AED} \asix{3F}{CBD}
    +\frac{1}{4} \lambda_{abcd} \asix{\phi F}{ABcd}\Bigg\},\\
    \asixd{\phi \widetilde{F}}{ABab} &= \sum_{\mathrm{perm}}\Bigg\{
 \afive{\phi F}{ACc} \afive{\phi \widetilde{F}}{BCd} \lambda_{abcd}
\nonumber \\ &\phantom{= \sum_{\mathrm{perm}}\Bigg\{}
+2g^2\theta^D_{bc}\theta^B_{cd}\afive{\phi F}{CAd}\afive{\phi\widetilde{F}}{CDa} + 10 g^2 \theta^D_{ac}\theta^D_{bd}\afive{\phi F}{CAc}\afive{\phi \widetilde{F}}{CBd}\nonumber\\
&\phantom{= \sum_{\mathrm{perm}}\Bigg\{}
-2g^2\theta^C_{ac}\theta^D_{bc}\afive{\phi F}{CAd}\afive{\phi\widetilde{F}}{DBd} + 6g^2 f^{CEF}f^{DEF}\afive{\phi F}{CAb}\afive{\phi\widetilde{F}}{DBa}
\nonumber\\
&\phantom{= \sum_{\mathrm{perm}}\Bigg\{}
-2\mathrm{i}g^2 f^{CDE}\theta^E_{bc}\afive{\phi F}{CAc}\afive{\phi\widetilde{F}}{DBa}
- 6g^2 f^{CEF}f^{DFB}\afive{\phi F}{CAb}\afive{\phi\widetilde{F}}{DEa}
\nonumber\\
&\phantom{= \sum_{\mathrm{perm}}\Bigg\{}
-\Big\{\mathrm{i}g\tr{t^B \afive{\psi F}{A}^\dagger \afive{\psi\phi^2}{ab}}
+2\mathrm{i}\tr{Y_a^\dagger \afive{\psi F}{A} Y_b^\dagger \afive{\psi F}{B}}
+\mathrm{h.c.}\Big\}
\nonumber \\ 
&\phantom{= \sum_{\mathrm{perm}}\Bigg\{}
-2\mathrm{i} g^3 \Big[ 
\asix{3\widetilde{F}}{ACD} \theta^B_{bd} \theta^C_{ac} \theta^D_{dc}
-\frac{3}{2} \asix{3\widetilde{F}}{ACD} \theta^B_{dc} \theta^C_{ac} \theta^D_{bd}
\Big]
\nonumber \\ 
&\phantom{= \sum_{\mathrm{perm}}\Bigg\{}
-2\mathrm{i}g \tr{
\asix{\psi F}{Ab} t^B Y_a^\dagger
- Y_a t^{B} \asix{\psi F}{Ab}^\dagger 
}
+\frac{1}{4} \asix{\phi \widetilde{F}}{ABcd} \lambda_{abcd}
\nonumber \\ 
&\phantom{= \sum_{\mathrm{perm}}\Bigg\{}
+\frac{1}{2}\theta^C_{ac}\theta^C_{bd} \asix{\phi\widetilde{F}}{ABcd}
-\theta^C_{ac}\theta^C_{cd} \asix{\phi\widetilde{F}}{ABdb}
\nonumber\\
&\phantom{= \sum_{\mathrm{perm}}\Bigg\{}
+2g^2\theta^B_{cd}\theta^C_{bd} \asix{\phi\widetilde{F}}{CAac}
-4g^2\theta^B_{bd}\theta^C_{cd} \asix{\phi\widetilde{F}}{CAac}\nonumber\\
&\phantom{= \sum_{\mathrm{perm}}\Bigg\{}
-\frac{11}{6}g^2 f^{ACD}f^{ECD}\asix{\phi\widetilde{F}}{EBab}
+\frac{1}{12}g^2 \theta^A_{cd}\theta^C_{dc}\asix{\phi\widetilde{F}}{CBab}\nonumber\\
&\phantom{= \sum_{\mathrm{perm}}\Bigg\{}
+\frac{1}{3}g^2 \tr{t^A t^C}\asix{\phi\widetilde{F}}{CBab}+\frac{1}{6} \tr{Y_c Y_a^\dagger+Y_a Y_c^\dagger} \asix{\phi\widetilde{F}}{ABcb}
\Bigg\},\\
    \asixd{\phi}{abcdef}&=\sum_{\mathrm{perm}}\Bigg\{ g^2\theta^{A}_{g c}\theta^{B}_{h b}\afive{\phi F} {A B f}\afive{\phi} {a h g d e} - 12 g^4\theta^{A}_{g d}\theta^{C}_{h c}\theta^{D}_{g b}\theta^{D}_{h a}\afive{\phi F}{A B f}\afive{\phi F} {C B e} \nonumber\\
    &\phantom{\sum_{\mathrm{perm}}\Bigg\{}- 6 g^4\theta^{A}_{g d}\theta^{B}_{h c}\theta^{C}_{g b}\theta^{D}_{h a}\afive{\phi F}{A B f}\afive{\phi F}{C D e} - \frac {5} {3} g^2\theta^{B}_{c g}\theta^{C}_{h g}\lambda_{a e f h}\afive{\phi F}{A C b}\afive{\phi F}{A B d}\nonumber\\
    &\phantom{\sum_{\mathrm{perm}}\Bigg\{}- \frac{2}{3} g^2\theta^{C}_{b h}\theta^{B}_{c h}\lambda_{a e f g}\afive{\phi F}{A B d}\afive{\phi F} {A C g}  + \frac{1}{3} g^2\theta^{B}_{ c g}\theta^{C}_{ h g}\lambda_{a ef h}\afive{\phi\tilde{F}}{A B b}\afive{\phi\tilde{F}} {A C d}\nonumber\\
    &\phantom{\sum_{\mathrm{perm}}\Bigg\{}- \frac{2}{3} g^2\theta^{C}_{b  h}\theta^{B}_{c  h}\lambda_{a e f g}\afive{\phi\tilde{F}}{A B g}\afive{\phi\tilde{F}}{A C d}  - \frac{1}{36} \lambda_{a b c g}\lambda_{d e f h}\afive{\phi F}{A  B  g}\afive{\phi F}{A  B  h}  \nonumber\\
    &\phantom{\sum_{\mathrm{perm}}\Bigg\{}- \frac{1}{36} \lambda_{a  b  c  g}\lambda_{d  e  f  h}\afive{\phi\tilde{F}}{A  B  g}\afive{\phi\tilde{F}}{A  B  h}  - \frac{1}{72} \afive{\phi}{a  e  f  g  h}\afive{\phi} {b  c  d  h  g}   \nonumber\\
    &\phantom{\sum_{\mathrm{perm}}\Bigg\{}
    +\tr{\afive{\psi \phi^2}{ab}^\dagger\afive{\psi \phi^2}{cd} Y_{e}^\dagger Y_f}
    +\Big\{\frac{1}{4}~\tr{
    \afive{\psi \phi^2}{ab}Y^\dagger_c \afive{\psi \phi^2}{de} Y^\dagger_f 
    }
    \nonumber\\
    &\phantom{\sum_{\mathrm{perm}}\Bigg\{}- \frac{1} {48} \tr{\afive{\psi\phi 2}{af}Y^{\dagger}_g}\afive{\phi}{g  b  c  d  e} 
    -20\lambda_{aef g} \tr{\afive{\psi\phi 2}{g d}\afive{\psi\phi2}{b c}^\dagger}
    \nonumber\\
    &\phantom{\sum_{\mathrm{perm}}\Bigg\{}+20\,\lambda_{a ef g}\tr{Y^{\dagger}_g\asix{\psi\phi}{bcd}} 
    -\frac{1}{3}~\tr{
    \asix{\psi \phi}{abc} Y^\dagger_d Y_e Y^\dagger_f
    }
    +\mathrm{h.c.}\Big\}\nonumber\\
    &\phantom{\sum_{\mathrm{perm}}\Bigg\{}+ 12 g^4\theta^{B}_{i  a}\theta^{A  h  c}\theta^{A}_{ g  d}\theta^{B}_{ h  b}\asix{\phi D} {g  e  i  f} - \frac{1}{2} g^2\theta^{A}_{g  h}\theta^{B}_{b  h}\lambda_{a  e  f  g}\asix{\phi F} {A  B  c  d}  \nonumber\\
    &\phantom{\sum_{\mathrm{perm}}\Bigg\{}- \frac{1}{8} g^2\theta^{A}_{h  a}\theta^{A}_{g  b}\asix{\phi}{c  d  e  f  h  g}  + 2 g^2\theta^{A}_{ b  g}\theta^{A}_{h  i}\lambda_{a  e  f  h}\asix{\phi D} {g  c  i  d} \nonumber\\
    &\phantom{\sum_{\mathrm{perm}}\Bigg\{}
    - \lambda_{a  d  g  h}\lambda_{b  c  h  i}\asix{\phi D}{i  e  g  f} - 3 g^4\theta^{A}_{g  d}\theta^{B}_{ h  c}\theta^{C}_{ g  b}\theta^{C}_{h  a}\asix{\phi F}{A  B  e  f}  \nonumber\\
    &\phantom{\sum_{\mathrm{perm}}\Bigg\{}- \frac{1}{6} \lambda_{a  e  f  g}\lambda_{g  b  h  i}\asix{\phi D}{h  i  c  d}  + \frac{1}{48} \lambda_{a  b  g  h}\asix{\phi}{c  d  e  f  h g} \Bigg\}.
\end{align}

In the process of reducing the bosonic redundant operators some fermionic operators are also generated, as can be seen from Section~\ref{sec_reduction}. These will be reported, together with the direct renormalization of fermionic operators in a companion paper~\cite{companion_fermions}.

\section{Conclusions\label{sec_conclusions}}

EFTs have become an essential tool in particle physics. The process of matching and running allows us to systematically incorporate the implications of arbitrary heavy physics into low energy experimental observables. While both matching and running calculations have been fully automated at the one-loop order~\cite{Carmona:2021xtq,Fuentes-Martin:2022jrf}, they have to be repeated on a model by model basis. A more efficient approach for finite matching calculations, in the case of the SMEFT, is based on the construction of IR/UV dictionaries~\cite{delAguila:2000rc,delAguila:2008pw,delAguila:2010mx,deBlas:2014mba,deBlas:2017xtg,Guedes:2023azv,Guedes:2024vuf}. In this article we have presented and alternative approach to optimize the calculation of the RGEs for arbitrary EFTs. Our approach is based on the definition of the most general EFT with an arbitrary number of scalar, fermion and gauge boson fields, with the flavor or symmetry structure of the EFT left completely general. We have defined the basis of local operators of such a general EFT up to mass dimension 6 both off-shell (Green's basis) and on-shell (physical basis), paying particular attention to the symmetry properties of the different operators and WCs. We have then computed the reduction of the Green's basis to the physical one, including non-linear contributions, paving this way their use in finite off-shell matching.~\footnote{Using the method recently described in~\cite{Chala:2024llp} for on-shell matching one could work directly with the physical basis.} 

Once the most general mass dimension 6 EFT has been defined, we have proceeded to compute the renormalization of the bosonic operators. This has been done using off-shell correlators, in the background gauge field, therefore obtaining the UV divergences in the Green's basis. Upon canonical normalization and reduction to the physical basis, we have obtained the beta function of all the bosonic operators in our physical basis. This result allows to compute, for any specific EFT, the beta function of any bosonic operator by means of a straight-forward group theoretic calculation. Computer tools, like \texttt{GroupMath}~\cite{Fonseca:2020vke}, can make this calculation trivial. All our results have been double-checked, either by using \texttt{MatchMakerEFT}~\cite{Carmona:2021xtq} and functional methods, or by using several different models to compare specific contributions. In particular we have performed partial cross-checks with the SMEFT beta functions~\cite{Jenkins:2013zja,Jenkins:2013wua,Alonso:2013hga}, full cross-check with the ALP-SMEFT~\cite{Chala:2020wvs,Bauer:2020jbp,Bonilla:2021ufe,DasBakshi:2023lca} as well as with many toy models.

The renormalization of fermionic operators, together with specific applications, will be published in a companion paper~\cite{companion_fermions}. Future generalization of our work can include the extension of this formalism to the two loop order, to mass dimension 8 or to finite matching. We plan to pursue these lines of research in the future.

\section*{Note added:} 

Some preliminary results of our calculation were presented by us (RF and JS) in SMEFT TOOLS 2022. At the same conference, M. Misiak and I. Nalecz presented some preliminary results along the same lines developed independently in collaboration with J. Aebischer, P. Mieszkalski and N. Selimovic. It is our understanding that some of these authors have continued working separately on the subject.

\section*{Acknowledgments}

We gratefully acknowledge very useful discussions with 
J. Aebischer, L. Bresciani, M. Chala, J.C. Criado, G. Guedes, J. Fuentes Martín, A. E. Thomsen and N. Selimovic and, in particular, we would like to thank M. Misiak and I. Nalecz for detailed comparisons on the possible bases of a general dimension six bosonic EFT.
This work has been partially supported by MCIN/AEI (10.13039/501100011033) and ERDF (grants PID2019-106087GB-C22 and PID2022-139466NB-C21), by the Junta de Andalucía grants FQM 101 and  P21\_00199 (FEDER) and by Consejería de Universidad, Investigación e Innovación, 
Gobierno de España and Unión Europea – NextGenerationEU under grant $\mathrm{AST22\_6.5}$, and by the European Union’s Horizon 2020 research and innovation programme under the
Marie Sklodowska-Curie grant agreement n. 101086085 – ASYMMETRY by the INFN Iniziativa Specifica APINE, and by the Italian MUR Departments of Excellence grant 2023-2027.

\appendix

\section{General calculation of the beta functions \label{appendix_RGE}}

The procedure to compute the beta functions of an EFT is well known, to the point of being textbook material by now. Nevertheless, we think that it is not easy to find a systematic, yet simple, description of the method in full generality. Mainly motivated by this and also for pedagogical reasons, we reproduce here the calculation of the beta functions for an arbitrary EFT at any order in perturbation theory. In part, the discussion below was inspired by the one in~\cite{Herren:2021yur}.

Let us define the beta function of a WC as follows
\begin{equation}
    \beta_{i} \equiv \mu \frac{\mathrm{d}a_i}{\mathrm{d}\mu},
\end{equation}
which can be computed from the fact that the bare WCs are independent of $\mu$. Indeed, we have
\begin{align}
\frac{\mathrm{d}a_i^{\mathrm{bare}}}{\mathrm{d}\ln \mu}=0
&= \mu^{(n_i-2)\epsilon} \left[ 
\epsilon (n_i -2) (a_i + \delta a_i) + \beta_i + \frac{\partial \delta a_i}{\partial a_j~}  \beta_j
\right] \nonumber \\
&= \mu^{(n_i-2)\epsilon} [ \epsilon (\overrightarrow{nC}+\overrightarrow{n\delta C} )+ (\mathbb{1}+M)\cdot \vec{\beta}], \label{cdotbare}
\end{align}
where we have used matrix notation in the last equality with the following definitions
\begin{align}
(\overrightarrow{na})_i \equiv & (n_i-2) a_i, \\
(\overrightarrow{n\delta a})_i \equiv & (n_i-2)\delta a_i, \\
M_{ij}\equiv & \frac{\partial \delta a_i}{\partial a_j~}, \\
(\vec{\beta})_i\equiv &  \beta_i.
\end{align}
From Eq. \eqref{cdotbare} we can write a formal expression for the beta functions, valid to any loop order,
\begin{equation}
\vec{\beta}=-\epsilon (\mathbb{1}+M)^{-1} \cdot(\overrightarrow{nC}+\overrightarrow{n\delta C} ).
\end{equation}
Note that in perturbation theory the inverse of the corresponding matrix is well defined in a perturbative sense.

Let us now consider the $\epsilon$ expansion of the above equation. The counterterms, and therefore the matrix $M$, have an expansion
\begin{equation}
\delta a_i = \sum_{n=0}^\infty \frac{a_i^{(n)}}{\epsilon^n},
\quad
M = \sum_{n=0}^\infty \frac{M^{(n)}}{\epsilon^n}. \label{countertermexpansion}
\end{equation}
It is therefore clear that the beta functions have a generic expansion of the form
\begin{equation}
\beta_i= \sum_{n=-1}^\infty \frac{\beta_i^{(n)}}{\epsilon^n}= \beta_i^{(-1)} \epsilon + \beta_i^{(0)} + \ldots~.
\end{equation}
The beta function in four dimensions is $\beta_i^{(0)}$ and more singular terms are normally considered to vanish, giving consistency conditions of the calculation at 2 and higher loops (see however~\cite{Herren:2021yur}). 
Introducing the expansions, Eq. \eqref{countertermexpansion} into the general expression for the beta function and keeping the finite term in the $1/\epsilon$ expansion we obtain the correct result. In order to get a closed expression let us consider the beta functions in the $\mathrm{MS}$ scheme (we can go from this to the $\mathrm{\overline{MS}}$ scheme by simply taking $\mu^2 \to \tilde{\mu}^2=4\pi \mathrm{e}^{-\gamma_E} \mu^2$). In that case we have $a_i^{0)}=M^{(0)}=0$ and therefore
\begin{equation}
    \vec{\beta}= - \epsilon \left[ 
    \overrightarrow{na}+ \frac{\overrightarrow{n\delta a}- M^{(1)}\cdot \overrightarrow{na}}{\epsilon} + \ldots
    \right]
    = - \epsilon~ \overrightarrow{na} 
    + M^{(1)}\cdot \overrightarrow{na}-\overrightarrow{n\delta a}
    +\mathcal{O}(1/\epsilon),
\end{equation}
so that $\beta_i^{(-1)} = - (n_i-2) a_i$ and,
\begin{equation}
    \beta_i^{(0)}=  (n_j-2) a_j 
\frac{\partial \delta a_i^{(1)}}{\partial a_j}    
- (n_i-2) \delta a_i^{(1)}.
\end{equation}
Finally, using the relation between the counterterms and the WCs of the divergent Lagrangian, Eq.~\eqref{deltCieqCiprime}, we obtain the final expression for the beta functions
\begin{equation}
\beta_i^{(0)} = \left[ 
(n_i-2) - (n_j-2) a_j \frac{\partial \phantom{a_j}}{\partial a_j}
\right] a_i^{\prime\,(1)}
= 
-2 L  a_i^{\prime\,(1)},
\end{equation}
where we have defined
\begin{equation}
    a_i^\prime = \sum_{n=1}^\infty \frac{a_i^{\prime\,(n)}}{\epsilon^n},
\end{equation}
and also the loop operator $L$ that, when acting on a perturbative expression gives, for each term in the expanded expression, the loop order of the term times the term itself. In order to obtain the last equality we have used that the divergence can be written as a sum of contributions over connected Feynman diagrams,
\begin{equation}
    a_i^{\prime\,(1)}=\sum_D  a_{i,D}^{\prime\,(1)},
\end{equation}
with $a_{i,D}^{\prime\,(1)}$ a polynomial function of WCs arising from diagram $D$, and that, for each connected Feynman diagram, the following topological identities hold
\begin{equation}
    L=P-V+1, \quad 2P+E=\sum_{v}n_v,
\end{equation}
where the sum in the second expression runs over all vertices in the diagram, $L$ stands for the number of loops, $P$ for the number of internal propagators, $V$ for the number of vertices, $n_v$ is the number of particles on vertex $v$ and $E$ is the number of external particles.
We then have
\begin{align}
    \left[ 
(n_i-2) - (n_j-2) a_j \frac{\partial \phantom{a_j}}{\partial a_j}
\right] a_{i,D}^{\prime\,(1)}
=&
\left[ E-2 - \sum_v (n_v-2)\right] a_{i,D}^{\prime\,(1)}
\nonumber \\
=&
\left[ E-2 - 2P-E+2V\right] a_{i,D}^{\prime\,(1)}
=-2L a_{i,D}^{\prime\,(1)}.
\end{align}
In the first identity we have used that $n_i=E$ and that
\begin{equation}
a_j \frac{\partial \phantom{a_j}}{\partial a_j},
\end{equation}
acting on $a_{i,D}^{\prime\,(1)}$ just runs over all vertices returning $a_{i,D}^{\prime\,(1)}$.

\section{UV poles in the Green basis \label{matching:green}}

For completeness we reproduce here the parameterization of the UV poles after canonical normalization but before reduction to the physical basis. In the process of these calculations, the computer code \texttt{SimTeEx}~\cite{Fonseca:2024rcg} has been extensively used to simplify the expressions.
\begin{align}
\ol
\eta^\prime_a &=
-\frac{1}{2} \kappa_{abc} (m_\phi^2)_{bc}
+ \tr{Y^a m_\psi^\dagger m_\psi m_\psi^\dagger 
+ \mathrm{h.c.}}
\nonumber \\ 
&-\frac{1}{2} \eta_b \Big\{
-2 g^2 \theta^A_{ac} \theta^A_{cb} + \frac{1}{2} \tr{
Y_a Y_b^\dagger +Y_b Y_a^\dagger}
-2(m_\phi^2)_{cd} \asix{\phi D}{abcd}
\Big\}
\\
\ol
(m_\phi^{2\,\prime})_{ab}&=
-\frac{1}{4} \kappa_{acd} \kappa_{bcd} 
-\frac{1}{4} \lambda_{abcd} (m_\phi^2)_{cd}
+\frac{3}{2}g^2 \theta^A_{ac} \theta^A_{cd} (m_\phi^2)_{db}
\nonumber \\ &
+\frac{1}{2}\tr{m_\psi^\dagger Y_a m_\psi^\dagger Y_b + \mathrm{h.c.}}
+2 \tr{m_\psi^\dagger Y_a Y_b^\dagger m_\psi}
-\frac{1}{4} (m_\phi^2)_{ac} \tr{Y_b Y_c^\dagger + \mathrm{h.c.}}
\nonumber \\ &
-\frac{1}{2} \Big[\tr{m_\psi^\dagger m_\psi m_\psi^\dagger \afive{\psi \phi^2}{ab} }+ \mathrm{h.c.}\Big]
\nonumber \\ &
+ \asix{\phi D}{abcd} (m_\phi^2)_{ce} (m_\phi^2)_{de}
+ \asix{\phi D}{acde} (m_\phi^2)_{bc} (m_{\phi^2})_{de}
+(a\leftrightarrow b),
\\
\ol
\kappa^\prime_{abc} &=
\sum_{\mathrm{perm}}
\Bigg\{
-\frac{1}{2}g^2 \Big[
\kappa_{ade} \theta^A_{bd} \theta^A_{ce}
+ \kappa_{abd} \theta^A_{ce} \theta^A_{de}
\Big] 
-\frac{1}{4} \lambda_{abde} \kappa_{cde}
\nonumber \\ & \phantom{\sum_{\mathrm{perm}}}
-\frac{1}{8} \kappa_{abd} \tr{
Y_a Y_d^\dagger + Y_a^\dagger Y_d}
+\tr{m_\psi Y_a^\dagger Y_b Y_c^\dagger
+m_\psi^\dagger Y_a Y_b^\dagger Y^c}
\nonumber \\ & \phantom{\sum_{\mathrm{perm}}}
+\frac{1}{12} \afive{\phi}{abcde} (m_\phi^2)_{de}
-\frac{1}{2} \Big\{\tr{\afive{\psi \phi^2}{ab}
(m_\psi^\dagger  Y_c m_\psi^\dagger
+2 m_\psi^\dagger m_\psi Y_c^\dagger)}
+ \mathrm{h.c.} \Big\}
\nonumber \\ & \phantom{\sum_{\mathrm{perm}}}
+2 \asix{\phi D}{abde} \kappa_{cdf} (m_\phi^2)_{fe}
+\frac{1}{3} \kappa_{abd} (m_\phi^2)_{ef} \Big[ 
\asix{\phi D}{cdef}-\asix{\phi D}{cedf}\Big]
\nonumber \\ & \phantom{\sum_{\mathrm{perm}}}
-\frac{1}{6} \Big\{
 \tr{m_\psi^\dagger m_\psi m_\psi^\dagger \asix{\psi \phi}{abc} }  + \mathrm{h.c.}\Big\} \Bigg\},
\\
\ol
\lambda^\prime_{abcd}&=
\sum_{\mathrm{perm}}
\Bigg\{
-\frac{3}{4} g^4 \theta^A_{ae} \theta^A_{bf} \theta^B_{ce}
\theta^B_{df}
-g^2 \Big[ 
\frac{1}{4} \lambda_{abef} \theta^A_{ce} \theta^A_{df}
+\frac{1}{6} \lambda_{abce} \theta^A_{df} \theta^A_{ef}
\Big]
-\frac{1}{16} \lambda_{abef} \lambda_{cdef}
\nonumber \\ & \phantom{\sum_{\mathrm{perm}}}
-\frac{1}{24} 
\lambda_{abce} \tr{Y_d Y_e^\dagger + Y_e Y_d^\dagger }
+\frac{1}{2} \tr{ Y_a Y_b^\dagger Y_c Y_d^\dagger}
\Big]
\nonumber \\ & \phantom{\sum_{\mathrm{perm}}}
+\frac{1}{12} \afive{\phi}{abcef} \kappa_{def}
-\Big\{\tr{\afive{\psi \phi^2}{ab} \Big(
\frac{1}{2} Y_c^\dagger m_\psi Y_d^\dagger 
+ Y_c^\dagger Y^d m_\psi^\dagger\Big)
} + \mathrm{h.c.}\Big\}
\nonumber \\ 
& \phantom{\sum_{\mathrm{perm}}}
+\frac{1}{2} 
\tr{ m_\psi^\dagger m_\psi \afive{\psi \phi^2}{cd}^\dagger \afive{\psi \phi^2}{ab}  }
+\frac{1}{8} \Big\{ 
\tr{\afive{\psi \phi^2}{ab} m_\psi^\dagger \afive{\psi \phi^2}{cd}
m_\psi^\dagger }
+\mathrm{h.c.} \Big\}
\nonumber \\ & \phantom{\sum_{\mathrm{perm}}}
+\frac{1}{48} \asix{\phi}{abcdef} (m_\phi^2)_{ef}
+\asix{\phi D}{abef} \kappa_{ceg} \kappa_{dfg} 
+\asix{\phi D}{abef} \lambda_{cdeg}(m_\phi^2)_{fg} 
\nonumber \\ & \phantom{\sum_{\mathrm{perm}}}
+\frac{1}{9}\Big(\asix{\phi D}{aefg} - \asix{\phi D}{afeg}\Big)
\lambda_{bcde} (m_\phi^2)_{fg}
-\frac{1}{6}\Big\{
\tr{
\asix{\psi \phi}{abc}\Big(
m_\psi^\dagger Y_d m_\psi^\dagger 
+ 2 m_\psi^\dagger m_\psi Y_d^\dagger
\Big)
}
+\mathrm{h.c.}\Big\}
\Bigg\},
\end{align}

Let us consider now the divergences parameterized by dimension-5 operators
\begin{align}
\ol
\afivep{\phi F}{ABa} &=
2 g^2 f^{ACD} f^{BED} \afive{\phi F}{ECa}
-\frac{g^2}{6} f^{CDE} f^{ADE} \afive{\phi F}{ACa} 
\nonumber \\ &
-g^2 \theta^{C}_{ac}\theta^B_{cb}\afive{\phi F}{ACb}
+\frac{g^2}{2} \theta^{C}_{ac}\theta^C_{cb}\afive{\phi F}{ABb}
-\frac{g^2}{12} \theta^{B}_{bc}\theta^C_{cb}\afive{\phi F}{ACa}
-\frac{g^2}{3} \tr{t^B t^C}
\afive{\phi F}{ACa}
\nonumber \\ &
-g \tr{t^B   \afive{\psi F}{A}^\dagger
Y_a
+Y_a^\dagger \afive{\psi F}{A}t^B}
-\frac{1}{8}
\tr{Y_a Y_b^\dagger + Y_b Y_a^\dagger}
\afive{\phi F}{ABb}
\nonumber \\ &
-\frac{1}{2}\kappa_{abc} \afive{\phi F}{ACb} \afive{\phi F}{BCc}
+\frac{1}{2}\kappa_{abc} \afive{\phi \widetilde{F}}{ACb} \afive{\phi \widetilde{F}}{BCc}
\nonumber \\ &
-\Big\{g \afive{\phi F}{ACa} 
\tr{\afive{\psi F}{B} t^C m_\psi^\dagger
+\afive{\psi F}{C} t^B m_\psi^\dagger
}+2\,\tr{
\afive{\psi F}{A}Y_a^\dagger \afive{\psi F}{B} m_\psi^\dagger
}+\mathrm{h.c.}\Big\}
\nonumber \\  
&-\frac{1}{4} \kappa_{abc} \asix{\phi F}{ABbc}
- g \Big\{\tr{\asix{\psi F}{Aa} t^B m_\psi^\dagger
}+\mathrm{h.c.}\Big\}
+(A\leftrightarrow B),
\\
\ol
\afivep{\phi \widetilde{F}}{ABa} &=
2g^2 f^{ACD}f^{BED} \afive{\phi \widetilde{F}}{ECa}
-\frac{g^2}{6} f^{CDE} f^{ADE} \afive{\phi \widetilde{F}}{ACa} 
\nonumber \\ &
-g^2 \theta^{C}_{ac}\theta^B_{cb}\afive{\phi \widetilde{F}}{ACb}
+\frac{g^2}{2} \theta^{C}_{ac}\theta^C_{cb}\afive{\phi \widetilde{F}}{ABb}
-\frac{g^2}{12} \theta^{B}_{bc}\theta^C_{cb}\afive{\phi \widetilde{F}}{ACa}
-\frac{g^2}{3} \tr{t^B t^C}
\afive{\phi \widetilde{F}}{ACa}
\nonumber \\ &
+\mathrm{i} g \tr{\afive{\psi F}{A} t^B Y_a^\dagger
-Y_a t^B \afive{\psi F}{A}^\dagger}
-\frac{1}{8} \afive{\phi \widetilde{F}}{ABb} 
\tr{
Y_a Y_b^\dagger+Y_b Y_a^\dagger
}
\nonumber \\ &
-\kappa_{abc} \afive{\phi F}{ACb} \afive{\phi \widetilde{F}}{BCc}
\nonumber \\ &
- \Big\{g\afive{\phi \widetilde{F}}{ACa} 
\tr{\afive{\psi F}{B} t^C m_\psi^\dagger
+\afive{\psi F}{C} t^B m_\psi^\dagger
}
-2\mathrm{i} \tr{
\afive{\psi F}{A}Y_a^\dagger \afive{\psi F}{B} m_\psi^\dagger
}
+\mathrm{h.c.}\Big\}
\nonumber \\ 
&
-\frac{1}{4} \kappa_{abc} \asix{\phi \widetilde{F}}{ABbc}
+\Big\{\mathrm{i} g \tr{\asix{\psi F}{Aa} t^B m_\psi^\dagger
}+\mathrm{h.c.}\Big\}
+(A\leftrightarrow B),
\\
\ol
\afivep{\phi}{abcde} &=
\sum_{\mathrm{perm}} \Big\{
\frac{1}{24}g^2 \theta^A_{fg} \theta^A_{ge} \afive{\phi}{abcdf}
-\frac{1}{12}g^2 \theta^A_{df} \theta^A_{eg} \afive{\phi}{abcfg}
-\frac{1}{24} \lambda_{defg} \afive{\phi}{abcfg}
\nonumber \\ &
-\frac{1}{96} \tr{Y_e Y_f^\dagger + Y_f Y_e^\dagger}
\afive{\phi}{abcdf}
+3 g^4 \theta^A_{eg} \theta^B_{df} \theta^C_{bf} \theta^C_{cg} \afive{\phi F}{ABa}
\nonumber \\ &
+\frac{1}{2} \tr{
\afive{\psi \phi^2}{ab} Y_c^\dagger Y_d Y_e^\dagger 
+\afive{\psi \phi^2}{ab}^\dagger Y_c Y_d^\dagger Y_e}
\nonumber \\ &
-\frac{1}{2}\tr{\afive{\psi \phi^2}{ab}\afive{\psi \phi^2}{cd}^\dagger Y_e m_\psi^\dagger
+\afive{\psi \phi^2}{ab}^\dagger \afive{\psi \phi^2}{cd}Y_e^\dagger m_\psi }
\nonumber \\ &
-\frac{1}{4}\tr{
\afive{\psi \phi^2}{ab}Y_c^\dagger\afive{\psi \phi^2}{de} m_\psi^\dagger
+\afive{\psi \phi^2}{ab}^\dagger Y_c \afive{\psi \phi^2}{de}^\dagger m_\psi
}
\nonumber \\ &
-\frac{1}{48} \asix{\phi}{abcdfg}\kappa_{efg}
-\asix{\phi D}{abfg} \lambda_{cdfh} \kappa_{egh}
\nonumber \\ &
+\frac{1}{6}\tr{
Y_a^\dagger \asix{\psi \phi}{bcd} Y_e^\dagger m_\psi
+Y_a \asix{\psi \phi}{bcd}^\dagger Y_e m_\psi^\dagger
}
\nonumber \\ &
+\frac{1}{3} \tr{
Y_a Y_b^\dagger \asix{\psi \phi}{cde} m^\dagger_\psi
+
Y_a^\dagger Y_b \asix{\psi \phi}{cde}^\dagger m_\psi
}
\Big\},
\\
\ol
\rfivep{\phi \Box}{abc} &=
3g^2 \theta^A_{ad} \theta^B_{db} \afive{\phi F}{ABc}
+\frac{1}{4} \tr{Y_a \afive{\psi \phi^2}{bc}^\dagger+Y_a^\dagger \afive{\psi \phi^2}{bc}}
\nonumber \\ &
+\asix{\phi D}{abde} \kappa_{cde}
-\frac{3}{2} \kappa_{ade} \asix{\phi D}{bcde}
+\mathrm{i} \tr{
\asix{\phi \psi}{ac} Y_b^\dagger m_\psi 
-m_\psi^\dagger Y_b \asix{\phi \psi}{ac}^\dagger
}
+(b \leftrightarrow c). \end{align}
And, finally, the ones corresponding to dimension 6 operators,
\begin{align}
\ol
\asixp{\phi}{abcdef}&=
\sum_{\mathrm{perm}}
\Big\{
\frac{1}{144}\afive{\phi}{abcgh}\afive{\phi}{defhg}
+6 g^4 \theta^A_{gb}\theta^C_{he}\theta^D_{ga}\theta^D_{hf}\afive{\phi F}{ABc} \afive{\phi F}{CBd}\nonumber\\
&\phantom{\sum_{\mathrm{perm}}+ \{}
+3 g^4 \theta^A_{gd} \theta^B_{hb}\theta^C_{ge}\theta^D_{ha}\afive{\phi F}{A Bc}\afive{\phi F}{CDf}
-\frac{1}{48}\theta^A_{ge}\theta^A_{hf}\asix{\phi}{abcdhg}\nonumber\\
&\phantom{\sum_{\mathrm{perm}}+ \{}
-\frac{1}{120}\theta^A_{ag}\theta^A_{hg}\asix{\phi}{bhcdef}
-\frac{1}{96}\lambda_{abgh}\asix{\phi}{cdefhg}\nonumber\\
&\phantom{\sum_{\mathrm{perm}}+ \{}
-6g^4\theta^A_{gc}\theta^A_{hb}\theta^B_{ha}\theta^B_{md}\asix{\phi D}{gemf}
+\frac{1}{2}\lambda_{abgh}\lambda_{cdhm}\asix{\phi D}{megf}
\nonumber\\
&\phantom{\sum_{\mathrm{perm}}+ \{}
+\frac{3}{2}\theta^A_{ge}\theta^B_{hb}\theta^C_{gf}\theta^C_{ha}\asix{\phi F}{ABcd}
\nonumber\\
&\phantom{\sum_{\mathrm{perm}}+ \{}
-\frac{1}{2}~\tr{\afive{\psi \phi^2}{ab}^\dagger\afive{\psi \phi^2}{cd} Y_{e}^\dagger Y_f}
-\frac{1}{8}~\tr{
\afive{\psi \phi^2}{ab}Y^\dagger_c \afive{\psi \phi^2}{de} Y^\dagger_f + \mathrm{h.c.}
}
\nonumber \\ &
\phantom{\sum_{\mathrm{perm}}+ \{}
+\frac{1}{6}~\tr{
\asix{\psi \phi}{abc} Y^\dagger_d Y_e Y^\dagger_f
+\mathrm{h.c.}
}
\Big\},\\
\ol
\asixp{3F}{ABC} &=
\sum_{\pi(ABC)} (-1)^{P_\pi} \Big\{
- g f^{DEA} \afive{\phi F}{DCa} \afive{\phi F}{EBa}
+ g f^{DEA} \afive{\phi \widetilde{F}}{DCa} \afive{\phi \widetilde{F}}{EBa}
\nonumber \\
&
\phantom{=
\sum_{\pi(ABC)} (-1)^{P_\pi} \Big\{}
- \frac{8}{9} \mathrm{i} \tr{\afive{\psi F}{B} t^C \overline{\afive{\psi F}{A}}}
-\frac{g^2}{12} \asix{3F}{ABD} f^{CEF} f^{DEF}
\nonumber \\ &
\phantom{=
\sum_{\pi(ABC)} (-1)^{P_\pi} \Big\{}
-\frac{1}{24} g^2 \asix{3F}{BCD}\tr{\theta^A \theta^D}
-\frac{1}{6} g^2 \asix{3F}{BCD}\tr{t^A t^D}
\Big\},
\\
\ol
\asixp{3\widetilde{F}}{ABC} &=
\sum_{\pi(ABC)} (-1)^{P_\pi} \Big\{
2 g f^{ADE} \afive{\phi F}{BDa}\afive{\phi \widetilde{F}}{CEa}
-\frac{g^2}{12} \asix{3\widetilde{F}}{ABD} f^{CEF} f^{DEF}
\nonumber \\ &
\phantom{
\sum_{\pi(ABC)} (-1)^{P_\pi} \Big\{}
-\frac{1}{24} g^2 \asix{3\widetilde{F}}{BCD}\tr{\theta^A \theta^D}
-\frac{1}{6} g^2 \asix{3\widetilde{F}}{BCD}\tr{t^A t^D}
\Big\},
\\
\ol
\asixp{\phi D}{abcd} =&\Big\{
\frac{1}{2}g^2 \Big[
\afive{\phi F}{ABa}\afive{\phi F}{ACb}\theta^B_{ce} \theta^C_{de}  
+\afive{\phi \widetilde{F}}{ABa}\afive{\phi \widetilde{F}}{ACb}\theta^B_{ce} \theta^C_{de} 
+(a \leftrightarrow b) 
\Big]
\nonumber \\ &
-\frac{1}{12} \tr{\afive{\psi \phi^2}{cd} \afive{\psi \phi^2}{ab}^\dagger}
-\frac{1}{2} \asix{\phi D}{abef} \lambda_{cdef}
\nonumber \\ &
+\frac{1}{6}g^2 \Big[
3 \theta^A_{bf} \theta^A_{ef} \asix{\phi D}{aecd}
- \Big( 
\theta^A_{bf} \theta^A_{ed} \asix{\phi D}{aecf}
+ (c \leftrightarrow d) \Big)
-7 \theta^A_{ae} \theta^A_{bf} \asix{\phi D}{efcd}
+(a\leftrightarrow b)\Big]
\nonumber \\ &
-\frac{\mathrm{i}}{6}\Big[
\tr{\asix{\phi \psi}{ac} Y_b^\dagger Y_d}
+(a \leftrightarrow b) + ( c \leftrightarrow d) + (ac \leftrightarrow bd)  \Big]
\nonumber \\  
&
-\frac{1}{3}\Big[-2g^2\theta^A_{af}\theta^A_{fe}\asix{\phi D}{ebcd} + \frac{1}{2}\tr{Y_a Y_e^\dagger+Y_e Y_a^\dagger}\asix{\phi D}{ebcd} +  (a\leftrightarrow b)\Big]
\nonumber\\
&
+(ab \leftrightarrow cd)
-\frac{1}{2} (a \leftrightarrow c)
-\frac{1}{2} (a \leftrightarrow d)
-\frac{1}{2} (b \leftrightarrow c)
-\frac{1}{2} (b \leftrightarrow d)
\Big\}
\\
\ol
\asixp{\phi F}{ABab} &= \Bigg\{
-\frac{1}{4} \afive{\phi F}{ACc} \afive{\phi F}{BCd} \lambda_{abcd}
\nonumber \\ &
-\frac{1}{2}g^2 \Big[ 
\theta^{D}_{ac} \theta^D_{bd} \afive{\phi F}{ACc} \afive{\phi F}{BCd}
-\theta^{C}_{ad} \theta^D_{bd} \afive{\phi F}{ACc} \afive{\phi F}{BDc}
-2\theta^{B}_{ad} \theta^D_{cd} \afive{\phi F}{ACc} \afive{\phi F}{CDb}
\Big]
\nonumber \\ &
+g^2 \Big[ 
3 f^{AEF} f^{BDF} \afive{\phi F}{CDb} \afive{\phi F}{CEa}
-5 f^{CEF} f^{BDF} \afive{\phi F}{ACb} \afive{\phi F}{DEa}
\nonumber \\ &
\phantom{g^2[}
+ f^{CEF} f^{DEF} \afive{\phi F}{ACb} \afive{\phi F}{BDa}
+ f^{ACE} f^{BDF} \afive{\phi F}{CDb} \afive{\phi F}{EFa}
\Big]
\nonumber \\ &
+\mathrm{i} g^2 \Big[ 
f^{ACE} \theta^D_{ac} \afive{\phi F}{CDb} \afive{\phi F}{BEc}
+f^{CDE} \theta^E_{bc} \afive{\phi F}{ACc} \afive{\phi F}{BDa}
\Big]
\nonumber \\ &
+\frac{1}{4} \afive{\phi \widetilde{F}}{ACc} \afive{\phi \widetilde{F}}{BCd} \lambda_{abcd}
\nonumber \\ &
+\frac{1}{2} g^2 \Big[ 
\theta^B_{bd} \theta^D_{cd} \afive{\phi \widetilde{F}}{ACc} \afive{\phi \widetilde{F}}{CDa}
-2 \theta^B_{cd} \theta^D_{bd} \afive{\phi \widetilde{F}}{ACc} \afive{\phi \widetilde{F}}{CDa}
\nonumber \\ &
\phantom{\frac{1}{2}g^2+}-\theta^B_{ad} \theta^D_{cd} \afive{\phi \widetilde{F}}{ACc} \afive{\phi \widetilde{F}}{CDb}
+\theta^D_{ac} \theta^D_{bd} \afive{\phi \widetilde{F}}{ACc} \afive{\phi \widetilde{F}}{BCd}
-\theta^C_{ad} \theta^D_{bd} \afive{\phi \widetilde{F}}{ACc} \afive{\phi \widetilde{F}}{BDc}
\Big]
\nonumber \\ &
+g^2 \Big[ 
2 f^{BDF} f^{CEF} \afive{\phi \widetilde{F}}{ACb} \afive{\phi \widetilde{F}}{DEa}
\nonumber \\ &
\phantom{+g^2}
-f^{AEF} f^{BDF} \afive{\phi \widetilde{F}}{CDb} \afive{\phi \widetilde{F}}{CEa}
-f^{CEF} f^{DEF} \afive{\phi \widetilde{F}}{ACb} \afive{\phi \widetilde{F}}{BDa}
\Big]
\nonumber \\ &
-\mathrm{i}g^2 \Big[f^{BCD} \theta^E_{bc} \afive{\phi \widetilde{F}}{ACc}
\afive{\phi \widetilde{F}}{DEa}
+f^{CDE} \theta^E_{bc} \afive{\phi \widetilde{F}}{ACc}
\afive{\phi \widetilde{F}}{BDa}
\Big]
\nonumber \\ &
+ \Big\{
\frac{1}{2}g\tr{t^B \afive{\psi F}{A}^\dagger \afive{\psi\phi^2}{ab}}
- \tr{\afive{\psi F}{A}Y_b^\dagger \afive{\psi F}{B} Y_a^\dagger}
+\mathrm{h.c.}\Big\}
\nonumber \\ &
-\frac{1}{4} \mathrm{i} g^3 \Big[ 
\theta^B_{ac} \theta^C_{cd} \theta^D_{db} 
-5 \theta^B_{bd} \theta^C_{ac} \theta^D_{cd}
+4 \theta^C_{ac} \theta^D_{bd} \theta^B_{cd}
\Big] \asix{3F}{ACD}
\nonumber \\ &
-\frac{1}{8} \asix{\phi F}{ABcd} \lambda_{abcd}
\nonumber \\ &
+\frac{1}{4} g^2 \Big[ 
-\theta^C_{ac} \theta^B_{bd} \asix{\phi F}{ABcd}
+2\theta^C_{cd} \theta^B_{bd} \asix{\phi F}{ACac}
+2\theta^C_{bd} \theta^B_{cd} \asix{\phi F}{ACac}
\Big] 
\nonumber \\ &
+g^2 \Big[ 
f^{ACE} f^{BDE} \asix{\phi F}{CDab}
-f^{CDE} f^{BDE} \asix{\phi F}{ACab}
\Big]
\nonumber \\ &
-\Big\{g \tr{\asix{\psi F}{Ab}t^B Y_a^\dagger}+\mathrm{h.c.}\Big\}
\nonumber \\ &
+(A\leftrightarrow B) + (a \leftrightarrow b) + (aA \leftrightarrow bB)\Bigg\},
\\
\ol
\asixp{\phi \widetilde{F}}{ABab} &= \Bigg\{
-\frac{1}{2} \afive{\phi F}{ACc} \afive{\phi \widetilde{F}}{BCd} \lambda_{abcd}
\nonumber \\ &
-g^2\theta^D_{bc}\theta^B_{cd}\afive{\phi F}{CAd}\afive{\phi\widetilde{F}}{CDa} - 5 g^2 \theta^D_{ac}\theta^D_{bd}\afive{\phi F}{CAc}\afive{\phi \widetilde{F}}{CBd}\nonumber\\
&+g^2\theta^C_{ac}\theta^D_{bc}\afive{\phi F}{CAd}\afive{\phi\widetilde{F}}{DBd} - 3g^2 f^{CEF}f^{DEF}\afive{\phi F}{CAb}\afive{\phi\widetilde{F}}{DBa}
\nonumber\\
&+\mathrm{i}g^2 f^{CDE}\theta^E_{bc}\afive{\phi F}{CAc}\afive{\phi\widetilde{F}}{DBa}
+ 3g^2 f^{CEF}f^{DFB}\afive{\phi F}{CAb}\afive{\phi\widetilde{F}}{DEa}
\nonumber\\
&+\Big\{\frac{\mathrm{i}}{2}g\tr{t^B \afive{\psi F}{A}^\dagger \afive{\psi\phi^2}{ab}}
+\mathrm{i}\tr{Y_a^\dagger \afive{\psi F}{A} Y_b^\dagger \afive{\psi F}{B}}
+\mathrm{h.c.}\Big\}
\nonumber \\ 
&
+\mathrm{i} g^3 \Big[ 
\asix{3\widetilde{F}}{ACD} \theta^B_{bd} \theta^C_{ac} \theta^D_{dc}
-\frac{3}{2} \asix{3\widetilde{F}}{ACD} \theta^B_{dc} \theta^C_{ac} \theta^D_{bd}
\Big]
\nonumber \\ &
+\mathrm{i}g \tr{
\asix{\psi F}{Ab} t^B Y_a^\dagger
- Y_a t^{B} \asix{\psi F}{Ab}^\dagger 
}
-\frac{1}{8} \asix{\phi \widetilde{F}}{ABcd} \lambda_{abcd}
\nonumber \\ &
-\frac{1}{4}\theta^C_{ac}\theta^C_{bd} \asix{\phi\widetilde{F}}{ABcd}
+\frac{1}{2}\theta^C_{ac}\theta^C_{cd} \asix{\phi\widetilde{F}}{ABdb}
\nonumber\\
&-g^2\theta^B_{cd}\theta^C_{bd} \asix{\phi\widetilde{F}}{CAac}
+2g^2\theta^B_{bd}\theta^C_{cd} \asix{\phi\widetilde{F}}{CAac}\nonumber\\
&+\frac{11}{12}g^2 f^{ACD}f^{ECD}\asix{\phi\widetilde{F}}{EBab}
-\frac{1}{24}g^2 \theta^A_{cd}\theta^C_{dc}\asix{\phi\widetilde{F}}{CBab}\nonumber\\
&-\frac{1}{6}g^2 \tr{t^A t^C}\asix{\phi\widetilde{F}}{CBab}-\frac{1}{12} \tr{Y_c Y_a^\dagger+Y_a Y_c^\dagger} \asix{\phi\widetilde{F}}{ABcb}
\nonumber\\&
+(A\leftrightarrow B) + (a \leftrightarrow b) + (aA \leftrightarrow bB)\Bigg\},
\\
\ol
\rsixp{2F}{AB}&=
-\frac{1}{3} \afive{\phi F}{A C a} \afive{\phi F}{B C a}
-\frac{1}{3} \afive{\phi \widetilde{F}}{A C a} \afive{\phi \widetilde{F}}{B C a}
-\frac{2}{3} \tr{\afive{\psi F}{A }^\dagger \afive{\psi F}{B }}
\nonumber \\ &
+\frac{1}{2} g \asix{3F}{A CD} f^{{BCD}} +(A\leftrightarrow B),
\\
\ol
\rsixp{FD\phi}{Aab} &=
\Bigg\{ 
 -\frac{17}{6} g f^{ABC} \Big[ 
\afive{\phi F}{DBa} \afive{\phi F}{DCb}
+\afive{\phi \widetilde{F}}{DBa} \afive{\phi \widetilde{F}}{DCb}
\Big]
\nonumber \\ &
+\frac{1}{2} g^2 \theta^B_{ac} \theta^C_{cb} \asix{3F}{ABC} 
+\frac{1}{3} \mathrm{i} g \theta^A_{cd} \asix{\phi D}{acbd}
+\frac{1}{3} g \tr{t^A \asix{\phi \psi}{ab}^T }
\Bigg\} - (a\leftrightarrow b),
\\
\ol
\rsixp{D\phi}{ab} &=
\afive{\phi F}{ABa} \afive{\phi F}{ABb}
+\afive{\phi \widetilde{F}}{ABa} \afive{\phi \widetilde{F}}{ABb}  
\\
\ol
\rsixp{\phi D}{abcd}= \sum_{(bcd)} &\Bigg\{ 
-2 g^2 \theta^B_{de} \theta^C_{cd} \afive{\phi F}{ABb} \afive{\phi F}{ACa}
-5 g^2 \theta^B_{de} \theta^C_{ad} \afive{\phi F}{ABb} \afive{\phi F}{ACc}
\nonumber \\ &
-2 g^2 \theta^B_{de} \theta^C_{cd} \afive{\phi \widetilde{F}}{ABb} \afive{\phi \widetilde{F}}{ACa}
+ g^2 \theta^B_{de} \theta^C_{ad} \afive{\phi \widetilde{F}}{ABb} \afive{\phi \widetilde{F}}{ACc}
\nonumber \\ &
-\frac{1}{12} 
\tr{\afive{\psi \phi^2}{ ab} \afive{\psi \phi^2}{cd}^\dagger
+\afive{\psi \phi^2}{ ab}^\dagger \afive{\psi \phi^2}{cd}
}
\nonumber \\ &
-\frac{3}{2} g^2 \Big[ 
-\theta^A_{de} \theta^A_{ef} \asix{\phi D}{afbc}
+\theta^A_{ce} \theta^A_{df} \asix{\phi D}{abef}
+\theta^A_{af} \theta^A_{be} \asix{\phi D}{efcd}
\Big]
-\frac{1}{2} \lambda_{abef} \asix{\phi D}{efcd}
\nonumber \\ &
+\frac{3}{2} g^2 \theta^A_{ae} \theta^B_{eb} \asix{\phi F}{ABcd}
+\frac{1}{12}
\tr{
Y_a \asix{\psi \phi}{bcd}^\dagger
+Y_a^\dagger \asix{\psi \phi}{bcd}
}
\Bigg\}. 
\end{align}

\bibliographystyle{style}
\bibliography{references}

\end{document}